\documentclass[usenatbib,twocolumn,12pt]{mn2e}
\usepackage{amsfonts}
\usepackage{amsmath}
\usepackage{amssymb}
\usepackage{graphicx}
\usepackage{color}
\usepackage{multirow}
\usepackage{booktabs}
\usepackage{tabu}
\usepackage{float}

\def\lsim{~\rlap{$<$}{\lower 1.0ex\hbox{$\sim$}}}
\def\bsim{~\rlap{$>$}{\lower 1.0ex\hbox{$\sim$}}}

\def\hmsun{\ {\rm M_\odot/{\it h}}}

\def\la{\langle}

\def\dd{{\rm d}}
\def\ln{{\rm ln}}

\def\dc{\delta_c}

\def\vk{\mathrm{\bf k}}

\def\fnl{f_\text{NL}}

\title[Scale-dependent bias]
      {Verifying the consistency relation for the scale-dependent bias from local primordial non-Gaussianity}

\author[M.Biagetti, T. Lazeyras, T. Baldauf, V. Desjacques, \& F. Schmidt]{\parbox{\textwidth}{
Matteo Biagetti$^{1,2}$, Titouan Lazeyras$^{3}$,Tobias Baldauf$^{4}$,   
Vincent Desjacques$^{1,5}$ and Fabian Schmidt$^{3}$\\}\\
$^1$ D\'epartement de Physique Th\'eorique and 
      Center for Astroparticle Physics (CAP), Universit\'e de Gen\`eve,      \\ ~~24 quai Ernest Ansermet, CH-1211 Gen\`eve, Switzerland \\
 $^2$ Institute of Physics, Universiteit van Amsterdam,
Science Park, 1098XH Amsterdam,  The Netherlands\\
$^3$ Max-Planck-Institut f\"ur Astrophysik, Karl-Schwarzschild-Str. 1, 85748 Garching, Germany\\
 $^4$ Department of Applied Mathematics and Theoretical Physics, University of Cambridge, \\
 ~~Wilberforce Road, CB3 0WA Cambridge, UK\\
 $^5$Physics department, Technion, Haifa 3200003, Israel \\}

\usepackage{xcolor}
\definecolor{MatteoColour}{rgb}{0.,0.7,0.3}
\definecolor{vincentColour}{rgb}{0.,0.05,0.9}
\definecolor{TitouanColour}{rgb}{0.7,0.05,0.7}
\definecolor{TobiasColour}{RGB}{242,101,34}

\begin{document}
\pagerange{\pageref{firstpage}--\pageref{lastpage}}

\maketitle 

\label{firstpage}

\begin{abstract}
We measure the large-scale bias of dark matter halos in simulations with non-Gaussian initial conditions of the local type, and compare this 
bias to the response of the mass function to a change in the primordial amplitude of fluctuations.  The two are found to be consistent, as 
expected from physical arguments, for three halo-finder algorithms which use different Spherical Overdensity (SO) and Friends-of-Friends 
(FoF) methods. On the other hand, we find that the commonly used prediction for universal mass functions, that the scale-dependent bias is 
proportional to the first-order Gaussian Lagrangian bias, does not yield a good agreement with the measurements.  
For all halo finders, high-mass halos show a non-Gaussian bias suppressed by 10--15\% relative to the universal mass function prediction.  
For SO halos, this deviation changes sign at low masses, where the non-Gaussian bias becomes larger than the universal prediction.
\end{abstract}


\section{Introduction}
\label{sec:intro}

Measurements of the anisotropies in the cosmic microwave background (CMB) point towards a Gaussian distribution of primordial fluctuations, 
with a nearly scale-invariant spectrum  (see \cite{Adam:2015rua} for the most recent results from the Planck satellite).
Nevertheless, testing the Gaussianity of the initial conditions is still an active research area as a detection of any departure from
Gaussianity could help discriminate among different classes of inflationary models which, as yet, predict initial conditions consistent with 
our current observations of the CMB (see \cite{Bartolo:2004if,Komatsu:2010hc, Chen:2010xka} for reviews).

For instance, single-field models, in which only one field is responsible for the generation of primordial perturbations, can generate a 
sizable three-point function, or bispectrum, for the equilateral model of primordial non-Gaussianity (PNG), where the three wave numbers are comparable (\cite{Babich:2004gb}).  
However, they  predict a negligible signal in squeezed configurations, which describes the coupling of large-scale modes with small-scale modes 
(\cite{Maldacena:2002vr}). By contrast, models with more than one field generate a sizable non-Gaussianity in squeezed configurations (e.g. the 
curvaton scenario, see \cite{Enqvist:2001zp,Lyth:2001nq, Moroi:2001ct}). 

Non-Gaussianity in the primordial curvature perturbations which peaks in squeezed configurations can be obtained with a simple parametrization
known as the \emph{local type} PNG. In this limit, the primordial gravitational potential is defined as
\begin{equation}
\Phi ({\bf x}) = 
\phi_{\rm G} ({\bf x}) + f_{\rm NL}^{\rm loc} (\phi^2_{\rm G} ({\bf x}) - \langle\phi^2_{\rm G}\rangle) + \mathcal{O}(\phi_{\rm G}^3)\;,
\end{equation}
where $f_{\rm NL}^{\rm loc} $ is the non-linearity parameter, $\phi_{\rm G}$ is a Gaussian field, and the last term indicates that the expansion 
can be extended to higher orders, which we will not consider in this analysis.

The Planck experiment has put the tightest constraints as of today on primordial non-Gaussianity, constraining $ f_{\rm NL}^{\rm loc}  = 0.8 \pm 5.0$ 
for the local type and $\fnl^{\rm equil} = -4 \pm 43$ for equilateral configurations (\cite{Ade:2015ava}). 
On the other hand, many multifield inflationary models predict $f_{\rm NL}^{\rm local}$ of order unity, and are consequently not excluded yet. Thus, improving constraints on local-type non-Gaussianity by an order of magnitude would be highly desirable.

While the current CMB limits are nearly cosmic-variance limited and, therefore, should not improve much in the future, observations of the large
scale structure in the late universe have the potential to outperform the current constraints.
Recent measurements of galaxy clustering and the integrated Sachs–Wolfe (ISW) effect are able to constrain $\fnl^{\rm loc}$ at the level of 
$\Delta\fnl \sim 30$  (\cite{Xia:2010pe}, \cite{Xia:2010yu, Xia:2011hj, Ross:2012sx,Giannantonio:2013uqa,Leistedt:2014zqa} ), while future galaxy 
redshift surveys are expected to yield constraints at least competitive with Planck (\cite{Giannantonio:2011ya,alonso/bull/etal:2015}).  
For instance, the Euclid mission has promising figures for galaxy power spectrum measurements, with $\Delta \fnl^{\rm loc} \sim 4$ 
(see \cite{Amendola:2012ys}). 
Even more promising, the combination of the galaxy power spectrum and bispectrum leads to a forecasted error of $\sigma(\fnl^{\rm loc}) = 0.2$  
for the NASA SPHEREx mission (\cite{Dore:2014cca}).
Future intensity mappings of the 21cm emission line of high-redshift galaxies should also interesting constraints with $\Delta \fnl^{\rm loc}\sim$ 
a few (\cite{Camera:2013kpa}). These errors could improve significantly if intensity maps are combined with galaxy redshift surveys 
(\cite{camera/santos/maartens:2015,fonseca/camera/etal:2015,alonso/bull/etal:2015}).

Primordial non-Gaussianity leaves various footprints in the formation of structures at late time (see \cite{Liguori:2010hx,Desjacques:2010jw} for 
reviews): the abundance of massive objects is enhanced (suppressed) for positive (negative) PNG; the clustering amplitude (bias) of 
galaxies relative to matter becomes strongly scale-dependent on large scales, and the 3-point function of galaxies encodes the shape of the 
primordial bispectrum (\cite{Nishimichi:2009fs,Sefusatti:2010ee,Baldauf:2010vn,Sefusatti:2011gt,Tellarini:2015faa,assassi/etal,Lazanu:2015rta}).  

In this analysis, we focus on this second signature, the \emph{scale-dependent bias} which was first noticed by \cite{Dalal:2007cu}  when measuring 
the cross halo-matter power spectrum in N-body simulations with non-Gaussian initial conditions of the local type. The large scale bias was found 
to have an additional, scale dependent, contribution 
\begin{align}
\frac{P_{\rm hm}}{P_{\rm mm}} =   b^{\rm G}_1  + \Delta b_\kappa(k,\fnl)\;,
\end{align} 
where we define
\begin{equation}\label{eq:deltabkappa}
\Delta b_\kappa(k,\fnl)= 2 \fnl \frac{b_{\rm NG}}{\mathcal M (k)} \;.
\end{equation}
They found that $b_{\rm NG}$ was proportional to the first-order bias\footnote{Here``univ'' stands for ``universal'', because this result is found to 
be valid for universal mass functions only, as we explain in the next paragraphs.},
\begin{equation}\label{eq:dalalnbody}
b_{\rm NG}^{\rm univ}=\dc b_1^{\rm L} \;.
\end{equation}
Here, $P_{\rm hm}$ is the cross halo-matter power spectrum, $P_{\rm mm}$ is the matter power spectrum, $\dc$ is the critical linear overdensity, 
$b_1^{\rm G}$ and $b_1^{\rm L}= b_1^{\rm G} -1$ are the Eulerian and Lagrangian bias, respectively, and
\begin{equation}
\mathcal{M}(k) = \frac 23 \frac{k^2 T(k) D(z)}{H^2_0 \Omega_{\rm m}} \;,
\end{equation} 
connects the linearly evolved matter density field and the primordial potential through the Poisson equation, where the transfer function $T(k)$ tends to unity at large scales, the linear growth rate $D(z)$ of density perturbations is normalised to 
$D(z) = 1/(1+z)$ during matter domination, and $H_0$ and $\Omega_{\rm m}$ are  the  present-day value  of  the  Hubble  rate  and  matter density, 
respectively. 
In  \cite{Matarrese:2008nc}, this signature was derived in the limit of high peaks, while \cite{Slosar:2008hx} showed that such a local-type 
non-Gaussianity in the primordial gravitational potential induces a local modulation of the amplitude of matter fluctuations proportional, at first 
order, to the non-linearity parameter $\fnl$. This modulation has an effect on the abundance of virialized halos, so that
\begin{equation}
\label{eq:pbs}
b_{\rm NG}^{\rm PBS}=\frac{\partial \ln \bar n_{\rm h}}{\partial \ln \sigma_8} \;.
\end{equation}
Here, ``PBS'' signifies ``peak-background split'', relating to the fact that the derivation of this behavior uses the separation of scales as it is 
usually done for the large scale bias of halos (\cite{Kaiser:1984sw,Bardeen:1985tr}).
This argument can be generalized to more general types of primordial non-Gaussianity (\cite{Schmidt:2010gw}).  
We express the mean halo overabundance $\bar{n}_h$  in terms of the number density of objects that have a mass in the interval $[M, M + dM]$, that is, 
the differential number density per unit volume and unit mass. 

Analytical models of the halo mass function (\cite{Press:1973iz,Bond:1990iw,Sheth:1999mn}) suggest that it is characterized by a first crossing 
distribution, or multiplicity function, $\nu f(\nu)$, 
\begin{equation}\label{eq:hmf}
\bar{n}_h(M,z)= \frac{M^2}{\bar\rho_m} \nu f(\nu) \frac{{\rm d}\ln M}{{\rm d}\ln \nu} \;,
\end{equation}
where $\nu(M,z) = \dc(z) / \sigma(M,z)$ is the peak height and $\sigma(M,z)$ is the amplitude of matter fluctuations for objects of mass $M$ at 
redshift $z$.
If the multiplicity function $f(\nu)$ depends only on the peak height $\nu$, the halo mass function is dubbed ``universal'' since all the redshift 
dependence is encoded in the peak significance $\nu$. In this case, and within the spherical collapse approximation, the non Gaussian bias amplitude 
is proportional to the first order Lagrangian bias (\cite{Slosar:2008hx,Ferraro:2012bd,PBSpaper,Scoccimarro:2011pz}), 
\begin{equation}\label{eq:euleriandb}
\frac{\partial \ln \bar n_{\rm h}}{\partial \ln \sigma_8} \stackrel{\bar n_{\rm h} {\rm \tiny{univ}}}{\longrightarrow} \dc b_1^{\rm L}(M) \;,
\end{equation}
such that Eq. \eqref{eq:dalalnbody}  and Eq. \eqref{eq:pbs} coincide in this limit.

The assumption of universality of the mass function has long been studied and its validity is still under debate 
(see \cite{Tinker:2008ff,Reed:2012ih,Despali:2015yla} and \cite{Pillepich:2008ka} for a discussion about universality in non-Gaussian simulations). 
Moreover, it is still unclear to which extent even a small deviation from universality may affect the non-Gaussian bias amplitude and therefore induce 
corrections in the relation of Eq. \eqref{eq:euleriandb}. 

\begin{table*}
\centering
\caption{Description of our 8 sets of simulations. In the last two columns, we quote numbers for the particle mass of each simulation and the minimum 
halo mass $M_\text{min}$ resolved. The latter corresponds to a minimum of $50$ particles per halo, below which a SO identification algorithm is not
reliable. }
\resizebox{0.65\textwidth}{!}{ 
\begin{tabular}{ccccccc}
\toprule
runs & N particles & L box (Gpc/h) & $\sigma_8$ & $ f_{\rm NL}^{\rm loc} $ &$M_{\rm part} (M_{\odot})$ & $M_{\rm min} (M_{\odot})$ \\
\hline\hline
4 & $1536^3$ & $2.0$ & $0.83$ & $0.0$ & $1.8 \times 10^{11} $ &$ 9.2 \times 10^{12} $  \\
6 & $1536^3$ & $2.0$ & $0.85$ & $0.0$ & $1.8 \times 10^{11}  $ &$ 9.2 \times 10^{12} $  \\
4 & $1536^3$ & $2.0$ & $0.87$ & $0.0$ & $1.8 \times 10^{11} $ &$ 9.2 \times 10^{12} $  \\
6 & $1536^3$ & $2.0$ & $0.85$ & $250.0$ & $1.8 \times 10^{11}  $ &$ 9.2 \times 10^{12} $  \\
6 & $1536^3$ & $2.0$ & $0.85$ & $-250.0$ & $1.8 \times 10^{11}  $ &$ 9.2 \times 10^{12} $  \\
1 & $1536^3$ & $1.0$ & $0.83$ & $0.0$ & $2.3 \times 10^{10}  $ &$ 1.1 \times 10^{12} $  \\
1 & $1536^3$ & $1.0$ & $0.85$ & $0.0$ & $2.3 \times 10^{10}  $ &$ 1.1 \times 10^{12} $  \\
1 & $1536^3$ & $1.0$ & $0.87$ & $0.0$ & $2.3 \times 10^{10} $ &$ 1.1 \times 10^{12} $  \\
\bottomrule
\end{tabular}}
\label{tab:sets}
\end{table*}

Previous analyses (see e.g.  \cite{Dalal:2007cu,Desjacques:2008vf,Pillepich:2008ka,Smith:2011ub}) always assumed the limit  of Eq. \eqref{eq:euleriandb} 
to be valid; with the important exception of \cite{Scoccimarro:2011pz} who, however, did not compute the modulation of the mass function relative to the 
local matter amplitude but to mass; and \cite{Matsubara:2012nc}, who considered generic Lagrangian bias models.
In agreement with an earlier analysis by \cite{Hamaus:2011dq}, they found some discrepancies between the measurement of the non Gaussian bias and 
the prediction from Eq. \eqref{eq:dalalnbody}, namely, the latter underestimates the effect of $\fnl$, when looking at halos identified with a 
Friends-of-Friends (FoF) algorithm. The same behaviour was confirmed recently by \cite{Baldauf:2015fbu}. 
By contrast, a similar analysis based on halos identified with a Spherical Overdensity (SO) finder found that Eq. \eqref{eq:dalalnbody} significantly 
overestimates the scale-dependent bias for halos with mass $M\sim M_\star$, see \cite{Hamaus:2011dq}.
Quantifying and understanding these discrepancies is particularly relevant for the forthcoming galaxy redshift surveys aiming at precise constraints 
of $\fnl$. 

The goal of this study is to accurately test the non-Gaussian bias correction in Eq. \eqref{eq:pbs}, ascertain the validity of the limit of Eq. 
\eqref{eq:euleriandb} and explore the sensitivity of our results to the particular choice of halo finder. We will measure the effect in N-body 
simulations that include dark matter (DM) particles solely. For this purpose, we adopt the following strategy:

\begin{enumerate}

\item Run 3 sets of simulations with Gaussian initial conditions and identical cosmologies, but for different values of the matter amplitude $\sigma_8$;

\item Run 2 sets of simulations with non-Gaussian initial conditions of the local type, with positive and negative values\footnote{Such values of $\fnl$ 
are, of course, excluded by current CMB constraints. Notwithstanding, since our focus is on the amplitude of the scale dependent bias proportional to $\fnl$ 
rather than the amplitude of $\fnl$ itself, we choose the largest value of $\fnl$ possible that is compatible with our linear treatment in $\fnl$, in 
order to get the cleanest possible signal.} of $\fnl=\pm 250$;

\item Estimate numerically the logarithmic derivative of the halo mass function $\bar n_{\rm h}$ w.r.t $\sigma_8$ using the 3 sets of simulations of point 
(i), and use another set of simulations with smaller volume to check the convergence of our measurements at low mass;

\item Measure the linear Eulerian bias $b_1^{\rm Eul} = 1+b_1^{\rm L}$ from the Gaussian simulations;

\item Measure the scale dependence of the halo power spectrum at large scales in the presence of primordial non-Gaussianity by estimating the cross 
halo-matter power spectrum $\langle \delta_{\rm h} \delta_{\rm m} \rangle / \langle \delta_{\rm m} \delta_{\rm m} \rangle$ in the non-Gaussian simulations 
under point (ii).

\end{enumerate}

Our paper is organized as follows. After introducing the details of our N-body simulations in \S\ref{sec:nbody}, we present our measurement in
\S\ref{sec:measurements} and discuss our results in \S\ref{sec:discussion}. We conclude in \S\ref{sec:conclusion}.

\section{The N-body simulations}
\label{sec:nbody}

Since our goal is to thoroughly investigate the scale dependence of halo bias at large scales in the presence of initial non-Gaussian conditions, 
on the one hand our simulations need to be run on a sufficiently large volume such that we can push for large scales and, on the other hand, they need 
to have high resolution to reliably identify individual halos and be able to trust the sensitivity down to low mass ranges.  
We achieve this goal by running the 8 sets of simulations outlined in Table \ref{tab:sets}. 

These simulations were run on the Baobab cluster at the University of Geneva and on the Odin cluster  at the Max Planck Institute in Garching. The 
cosmology is a flat $\Lambda$CDM model with $\Omega_{\rm m}=0.3$, $h=0.7$, $n_s=0.967$ and varying $\sigma_8$ as shown on Table \ref{tab:sets}. The transfer 
function was obtained from the Boltzmann code CLASS (\cite{Blas:2011rf}).  The initial particle displacements were implemented at $z_i= 99$ using the 
public code 2LPTic (\cite{Scoccimarro:1997gr,Crocce:2006ve}) for realizations with Gaussian initial conditions and its modified version 
(\cite{Scoccimarro:2011pz}) for non-Gaussian initial conditions of the local type. The simulations were evolved using the public code Gadget2 
(\cite{Springel:2005mi}). 

We perform our measurement using three different algorithms for finding DM halos. We consider the Spherical Overdensity (SO) algorithm Amiga Halo Finder 
({\small AHF})
(\cite{Gill:2004km,Knollmann:2009pb}), using a redshift-independent overdensity of $\Delta=200$ with respect to the background matter density. 
The first of the two different Friends-of-Friends (FoF) finders considered is {\small Rockstar} 
(\cite{Behroozi:2011ju}), for which we use a linking length of $\lambda=0.28$. Since Rockstar uses an FoF algorithm to find halos, but estimates the halo 
mass with a SO approach, we shall refer to it as ``Hybrid''. The code provides several prescriptions to measure the SO halo mass, we choose to be consistent with the AHF prescription and use again $\Delta=200 \bar\rho_m$.  Finally, we employ  a genuine Friends-of-Friends algorithm with two different linking lengths of $\lambda=0.15$ and $\lambda=0.2$, which we shall simply refer 
to as {\small FoF}.  

\section{Theory}\label{sec:theory}
The quantity we want to measure is the scale-dependent shift, $\Delta b_\kappa(k,\fnl)$, introduced in Eq.\eqref{eq:deltabkappa}, to the ratio between 
the halo-matter cross power spectrum in non-Gaussian simulations over the matter auto power spectrum in Gaussian simulations. We model this quantity, following  \cite{Desjacques:2008vf}, as
\begin{align}
\label{eq:deltabkhm}
\frac{P^{\rm NG}_{\rm hm}(k,\fnl)}{P^{\rm G}_{\rm mm}(k,0)} & = b^{\rm G}_{\rm hm} + \Delta b_I(\fnl) + b^{\rm G}_{\rm hm} \beta_m(k,\fnl) \nonumber \\
&+ \Delta b_{\kappa}(k,\fnl) + \mathcal{O}(b^{\rm G}_2, \fnl^2).
\end{align}
This formula describes only the leading order corrections, where by leading order here we mean at first order both in the bias and in the nonlinear parameter
$\fnl$ and higher order terms are included in the term $\mathcal{O}(b^{\rm G}_2, \fnl^2)$. Since our focus is on the effect at the largest scales, higher order corrections will not be analyzed in detail, as they only become relevant on small scales. We will however comment on them at the end of this section. 
In addition to the linear Gaussian halo bias $b^\text{G}_\text{hm}$ measured as in  Eq. \eqref{eq:bhm} and the scale-dependent bias $\Delta b_\kappa(k,\fnl)$, which dominates at low wavenumber, we have taken into account two additional contributions.

Firstly, there is a scale-independent correction
\begin{equation}
\Delta b_I(\fnl)=-\frac{1}{\sigma(M)} \frac{\partial}{\partial \nu} \ln \left[ \frac{f(\nu,\fnl)}{f(\nu,0)}\right] ,
\end{equation}
which arises from the change in the mean number density of halos (hence the slope of the mass function) in the presence of PNG 
\citep[see][]{Afshordi:2008ru,Desjacques:2008vf}. This effect grows  with increasing halo mass, given that it is inversely proportional to the variance $\sigma(M)$ and that the effect of $\fnl$ peaks at the high mass tale of the halo mass function. Also, the correction has opposite sign with respect to $\fnl$, since the bias decreases (increases) whenever the halo mass function is enhanced (suppressed), as in the case of a positive (negative) $\fnl$.

Secondly, the matter power spectrum also changes in the presence of PNG \citep{Scoccimarro:2003wn,Grossi:2008fm,Taruya:2008pg, Pillepich:2008ka}, and
this induces a correction of the form
\begin{equation}
\beta_m(k,\fnl) = \frac{P_{\rm mm}(k,\fnl)-P_{\rm mm}(k,0)}{P_{\rm mm}(k,0)}.
\end{equation}
Here $P_{\rm mm}(k,0)$ and $P_{\rm mm}(k,\fnl)$ are the matter power spectrum from the Gaussian and non-Gaussian initial conditions, respectively. Being 
a loop correction, this term vanishes on large scales and becomes more important with increasing wavenumber, and is thus qualitatively different 
from that of $\Delta b_\kappa(k,\fnl)$.

 Notice that, for the high values $\fnl = \pm 250$ adopted here, second order effects proportional to $\fnl^2$ may be important. 
However, we can take advantage of our simulations with both negative and positive $\fnl$ and cancel contributions up to order $\mathcal{O}(\fnl^3)$, as we will explain in the next section.

  In our expression for the halo-matter power spectrum, we have only included the leading terms relevant on large scales, which correspond to linear bias operators. All higher order, nonlinear bias terms only enter at the loop level, and are suppressed by powers of  $(k/k_{\rm NL})^{3+n}$, where $k_{\rm NL}$ is the nonlinear scale, and $k_{\rm NL}  \approx 0.3 h {\rm Mpc}^{-1}$ at redshift zero, while $n \approx -1.5$ is the index of the matter power spectrum on the scales of interest. Since we restrict our fits to scales of $k < 0.03 h {\rm Mpc}^{-1}$, neglecting these terms does not bias our estimate of $b_{\rm NG}$.

Note also that, as shown in \cite{McDonald:2008sc,assassi/etal}, the large-scale scale-dependent bias proportional to $b_2$ first proposed in \cite{Matarrese:2008nc} is renormalized into the parameter $b_{\rm NG}$ in the general perturbative bias expansion; that is, there is no additional contribution $\propto b_2$ on large scales.

\section{Measurements}
\label{sec:measurements}

In this section, we provide details of our measurements on the N-body simulations with the specifications given above. Here and henceforth, error bars 
represent the standard deviation of the mean calculated from the different realizations, 
\begin{equation}
\sigma_{mean}=\sqrt{\frac{\sum_{i=1}^N (x_i - \mu)^2}{N(N-1)}} \;,
\end{equation}
where $x_i$ is the value for the i-th realization, $N$ is the number of realizations and $\mu$ is the mean among the realizations.

\subsection{Halo mass function}

We measure the halo mass function for each set of simulations for the three algorithms by counting halos in logarithmically spaced mass bins. 
 We ran the halo finders on the outputs at redshift $z=0$, $1$ and $2$ in order to explore the redshift dependence of our results. 

We compare the measurement for the SO and Hybrid halos to the fitting formulae of \cite{Tinker:2010my} (hereafter Ti10) and, for the FoF halos, 
to that of \cite{Sheth:1999mn} (hereafter S\&T99). Note that the mass function of S\&T99 is of the form of Eq. \eqref{eq:hmf}, with a multiplicity 
function given by 
\begin{equation}
\nu^2 f(\nu) = A \left(1+\frac{1}{\nu'^{2p}}\right)\left(\frac{\nu'^2}{2}\right)^{1/2} \frac{e^{-\nu'^2/2}}{\sqrt{\pi}}
\end{equation} 
where $\nu'=\sqrt{q}\nu$. We use the fitting values $A=0.322$ and $0.368$, $p=0.3$ and $0.25$ and $q=0.8$ and $0.7$ for $\lambda=0.15$ and $0.2$, 
respectively.
 
\begin{figure*}
\centering 
\resizebox{0.33\textwidth}{!}{\includegraphics{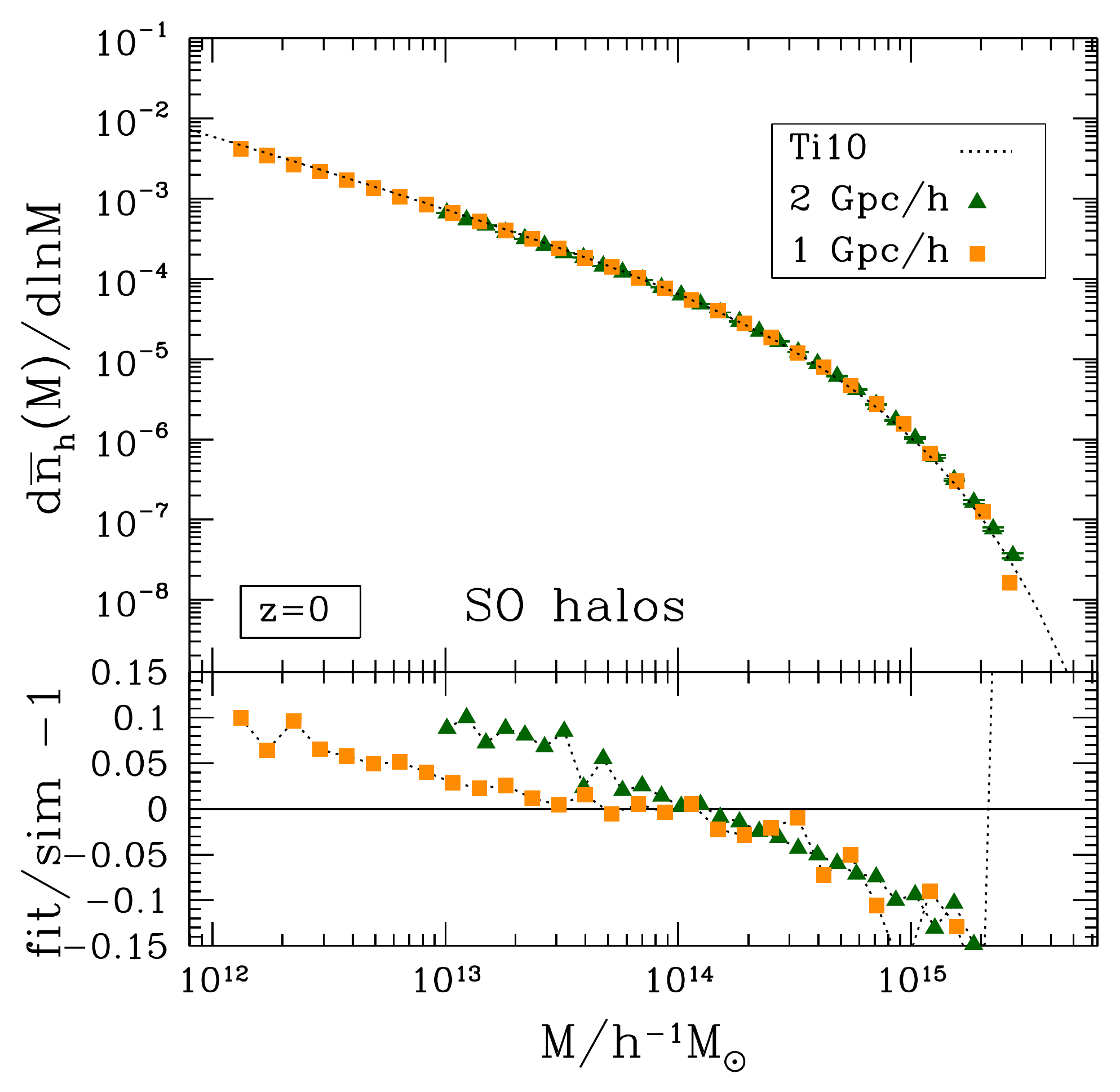}}
\resizebox{0.33\textwidth}{!}{\includegraphics{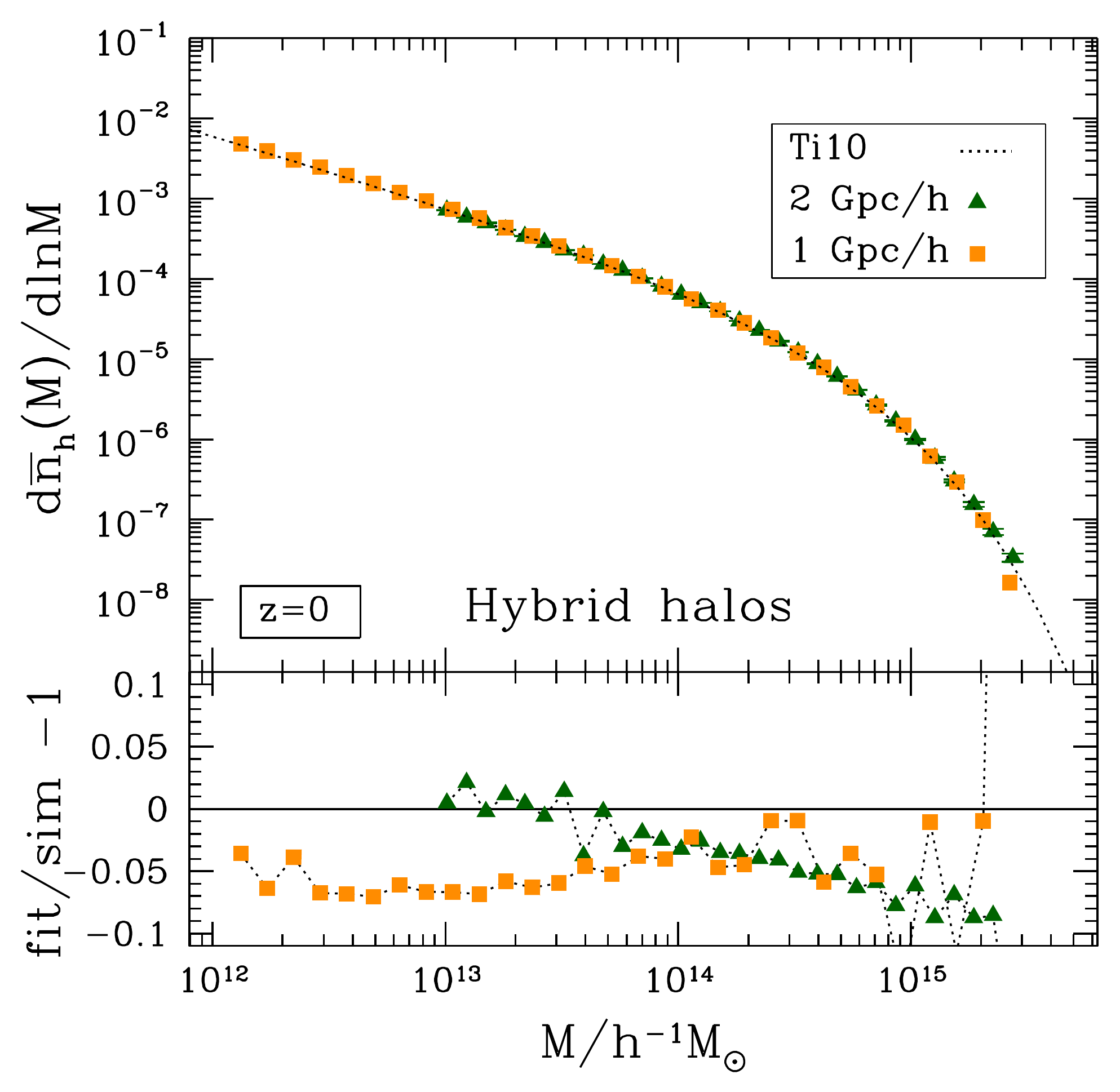}}
\resizebox{0.33\textwidth}{!}{\includegraphics{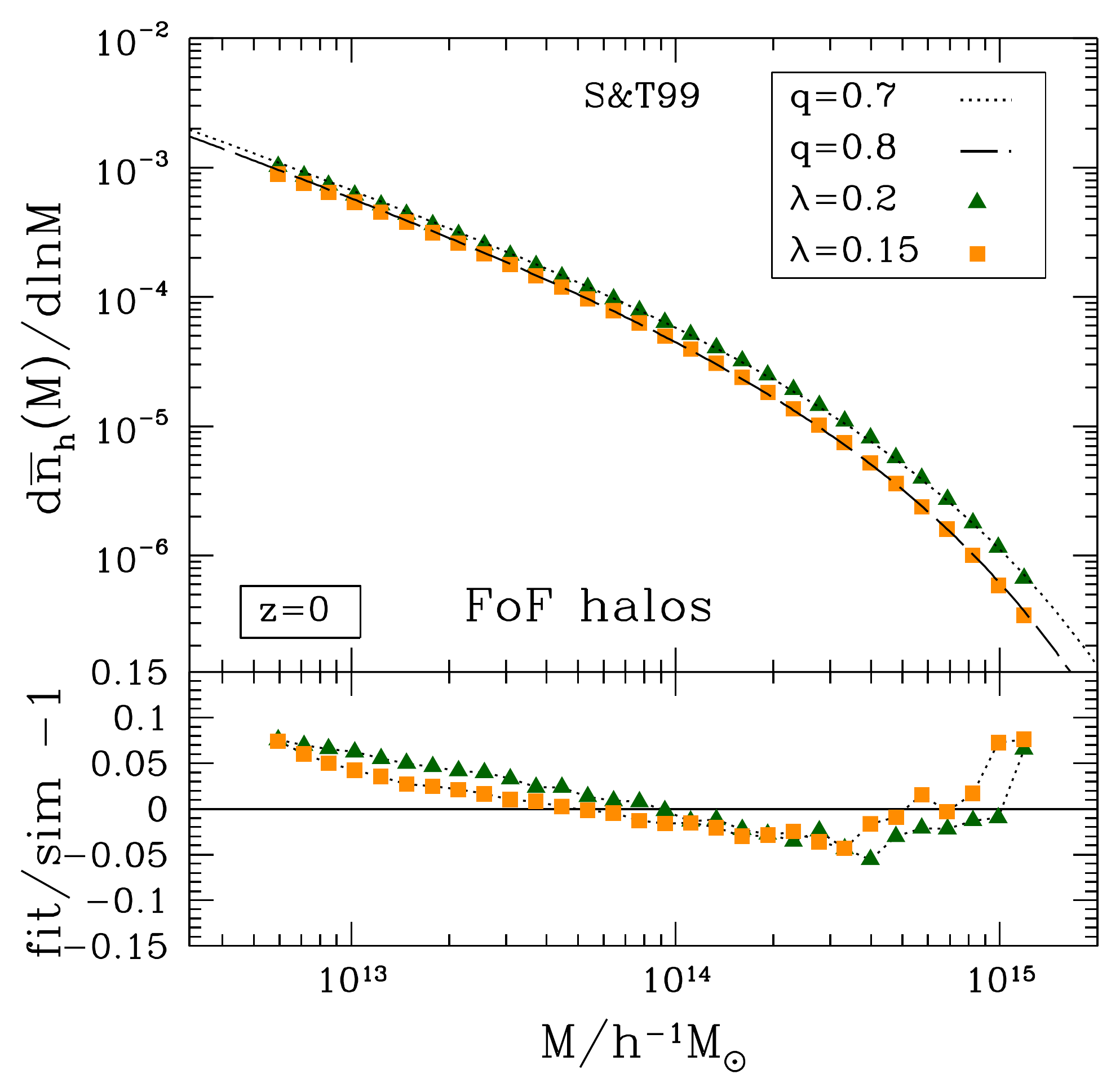}}
\caption{Halo mass function for the Gaussian simulations with $\sigma_8=0.85$, where both box sizes for the SO  (left) and Hybrid (center) algorithms are included. In the right panel we show the FoF halos for the $2$ Gpc/h box and for two different values of the linking length. Corresponding fits are shown as dotted and long dashed lines, respectively. In the lower panels, we show the relative difference between the fits and the measurements. }
\label{fig:ahf}
\end{figure*}

In Figure \ref{fig:ahf}, we show the results for the Gaussian simulations with $\sigma_8=0.85$. For the SO and Hybrid halo finders, we consider the two 
box sizes in order to assess the convergence at low mass whereas, for the FoF halos, we measure the mass function from the 2 Gpc/h boxes using two 
different linking lengths, $\lambda=0.15$ and 0.2.

\subsection{Linear bias}

For Gaussian initial conditions and at sufficiently large, i.e. linear, scales, the halo bias is scale independent, but dependent on the mass and 
redshift of the halo population considered. This constant value can be measured by taking the ratio of the halo and matter power spectra
\begin{equation}
b_{\rm hm}^{\rm G} = \frac{P^{\rm G} _{\rm hm}}{P^{\rm G} _{\rm mm}}.\label{eq:bhm}
\end{equation}
To measure these power spectra, we extract the dark matter and halo fluctuation fields $\delta_{\rm m}(k)$ and $\delta_{\rm h}(k)$ 
by interpolating particles (dark matter and halo centers) on a three dimensional grid of size $512^3$. 

Notice that the linear bias can be computed also using the ratio 
\begin{equation}
b_{\rm hh}^{\rm G}  = \left(\frac{P^{\rm G} _{\rm hh}-C}{P^{\rm G} _{\rm mm}}\right)^{1/2} \,,
\end{equation}
where in this case one needs to substract the shot noise $C$. Recent studies  indicate that the shot noise may deviate from the constant value
$C=1/\bar n$ which is assumed for DM halos that are Poisson sampled (see  \cite{Baldauf:2013hka, Baldauf:2015fbu} and Appendix \ref{app:stoca} for more 
discussion about this). We therefore consider the cross value $b_{\rm hm}$ only for the present analysis.

We chose to split the halo catalogs into three mass bins with equal number of halos for the $2$ Gpc/h box simulations, adding two bins at lower 
mass for the smaller, $1$ Gpc/h box. The characteristics of these halo bins are displayed in Table \ref{tab:bias}, along with the values of the 
corresponding linear halo bias. We measured the latter upon taking ratios as in \eqref{eq:bhm} and averaging over the wavenumber interval 
$k \in [0.004,0.03]$ h/Mpc since, at higher wavenumbers, higher order biases (such as $b_2$) start to contribute significantly.
Note also that we define the central mass value of each bin to be
\begin{equation}\label{eq:mean}
\bar M = \frac{\int_{M_\text{min}}^{M_\text{max}} \dd M\, M\, \bar n_h(M)}{\int_{M_\text{min}}^{M_\text{max}} \dd M\, \bar n_h(M)} \;,
\end{equation}
where $\bar n_h$ is the halo mass function fit, Ti10 in case of SO/Hybrid halos and S\&T99 in case of FoF halos.
A plot of all the measured ratios as in \eqref{eq:bhm} are displayed in Appendix, Figs. \ref{fig:lbias}.
\begin{center}
\begin{table*}
\begin{tabular}{ |c|c|c|c|c|c|c|c|c|c|c|c|c|c| }
\hline
$\mbox{}$&$\mbox{}$&$\mbox{}$  &  SO  & Hybrid  &$\mbox{}$ & SO  & $\mbox{}$ & Hybrid  &$\mbox{}$ & FoF &  ($\lambda=0.2$)   & FoF & ($\lambda=0.15$)  \\
\hline
 Mass & $\bar M$ & {\small $1$ Gpc/h}&$b_{\rm mh}$ &  $b_{\rm mh}$ & {\small $2$ Gpc/h}& $b_{\rm mh}$ &  $\sigma_{b_{\rm mh}}$& $b_{\rm mh}$ &  $\sigma_{b_{\rm mh}}$ & 
$b_{\rm mh}$ &  $\sigma_{b_{\rm mh}}$&$b_{\rm mh}$ &  $\sigma_{b_{\rm mh}}$  \\
\hline\hline
$1.1 - 2.2$ & $1.6$ &$\mbox{}$& $0.74$ &  $0.88$ &$\mbox{}$  &$-$ & $-$ & $-$ & $-$ & $-$ & $-$ & $-$ & $-$  \\
$2.2 - 9.2$ & $4.2$  &$\mbox{}$&$0.77$  & $0.92$&$\mbox{}$   & $-$ & $-$ & $-$ & $-$ &  $-$ & $-$ & $-$ & $-$  \\
$9.2 - 14$ & $11.1$ &$\mbox{}$& $1.07$ &  $1.20$ &$\mbox{}$  &$1.01$ & $0.02$ & $1.13$ & $0.01$ & $1.05$ & $<0.01$ & $1.11$ & $<0.01$  \\
$14 - 27$ & $18.9$  &$\mbox{}$&$1.18$  & $1.26$&$\mbox{}$   & $1.14$ & $0.01$ & $1.27$ & $0.01$ &  $1.13$ & $<0.01$ & $1.22$ & $<0.01$  \\
$27 - 3000$ &$82.9$&$\mbox{}$& $1.56$ & $1.67$&$\mbox{}$  & $1.65$ & $0.01$ &  $1.73$ & $0.01$ & $1.62$ & $<0.01$ & $1.73$ & $<0.01$ \\
\hline
\end{tabular}
\caption{{ Measured values for linear bias at redshift $z=0$ for the Gaussian simulations where mass ranges are expressed in units of $10^{12} M_{\odot}$ 
and refer to the $2$Gpc/h box sets  and the $1$Gpc/h box sets.}}
\label{tab:bias}
\end{table*}
\end{center}

\subsection{Scale dependent bias}

\begin{figure*}
\centering 
\resizebox{0.43\textwidth}{!}{\includegraphics{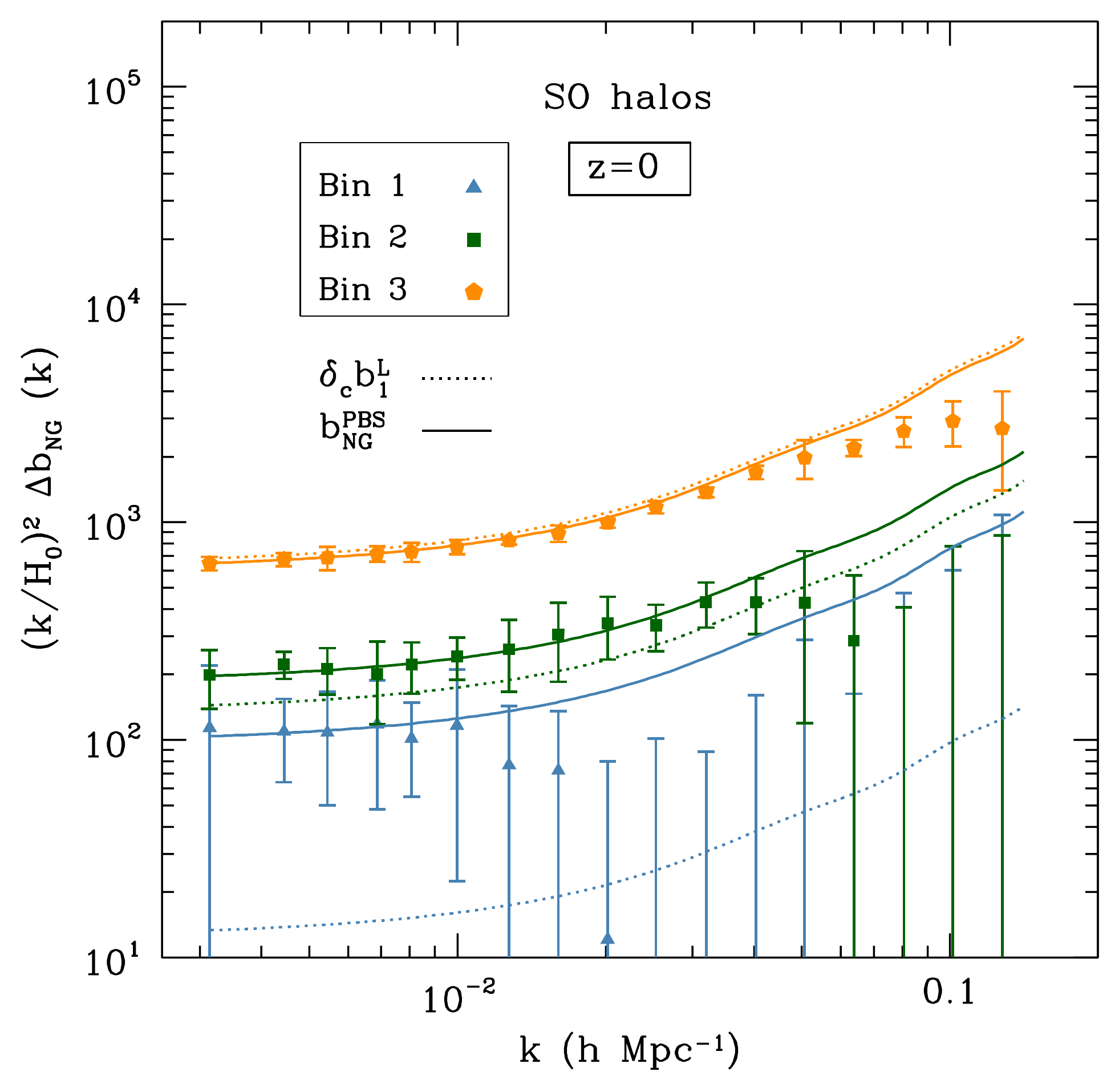}}
\resizebox{0.43\textwidth}{!}{\includegraphics{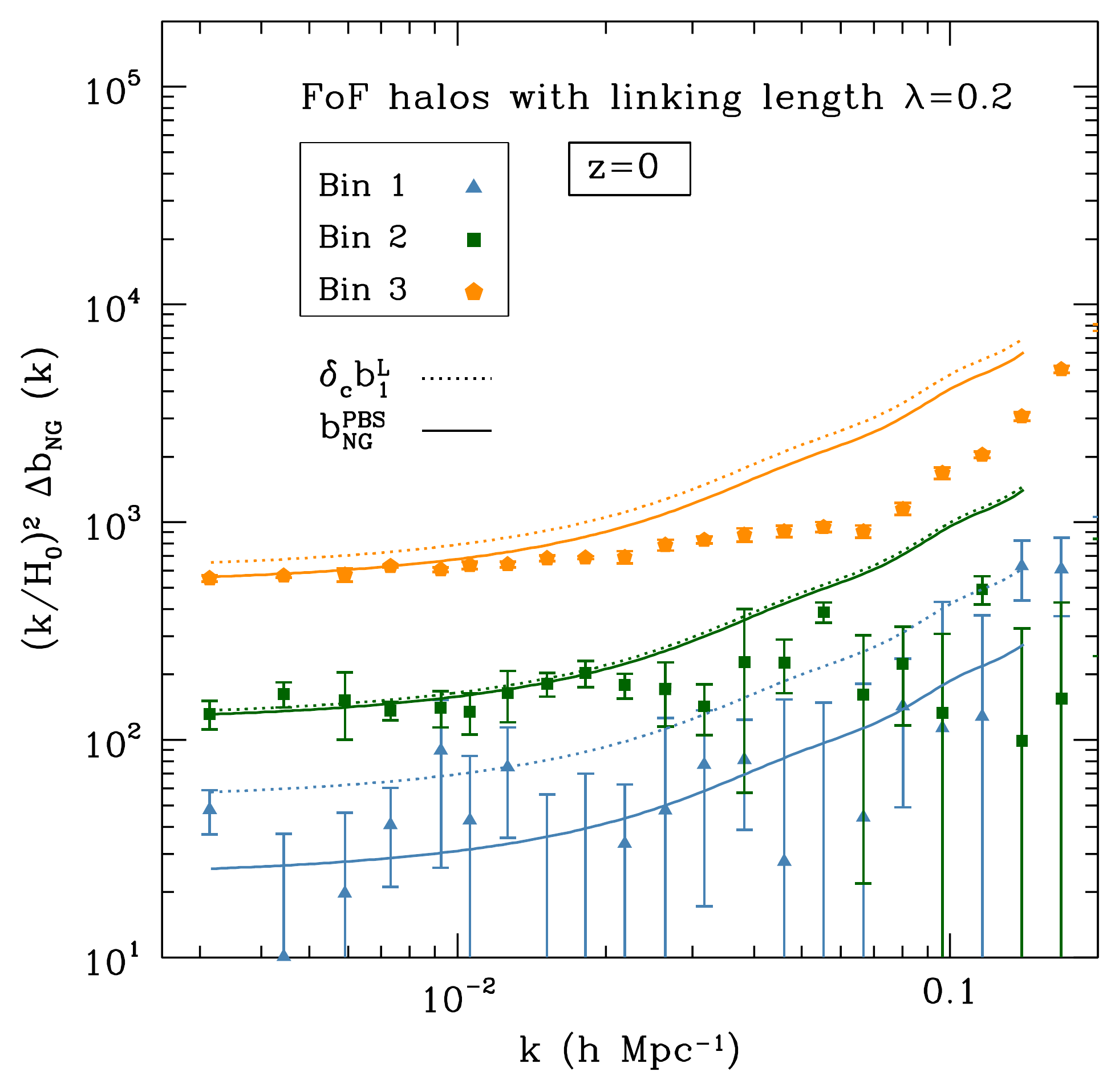}}
\resizebox{0.43\textwidth}{!}{\includegraphics{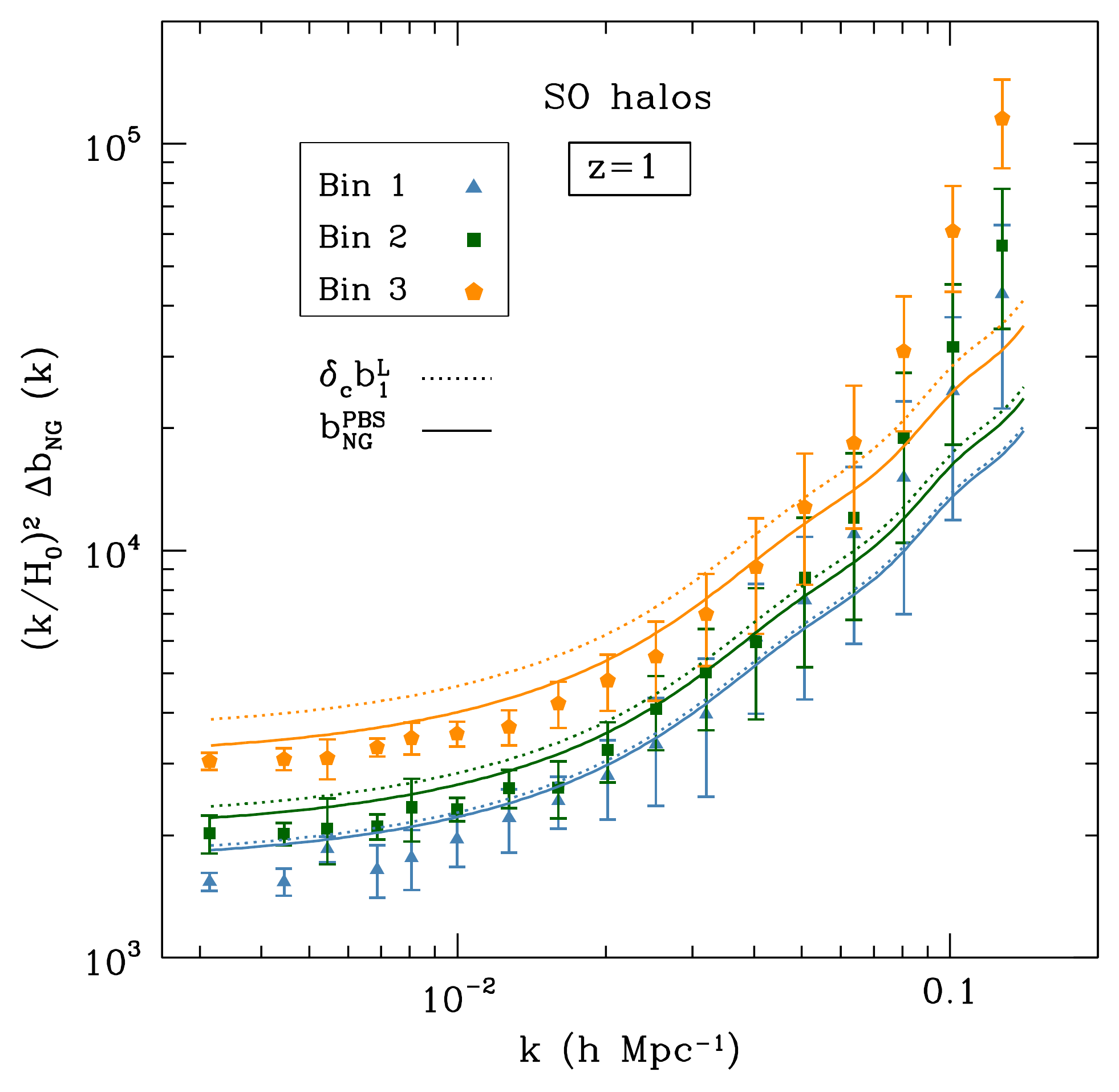}}
\resizebox{0.43\textwidth}{!}{\includegraphics{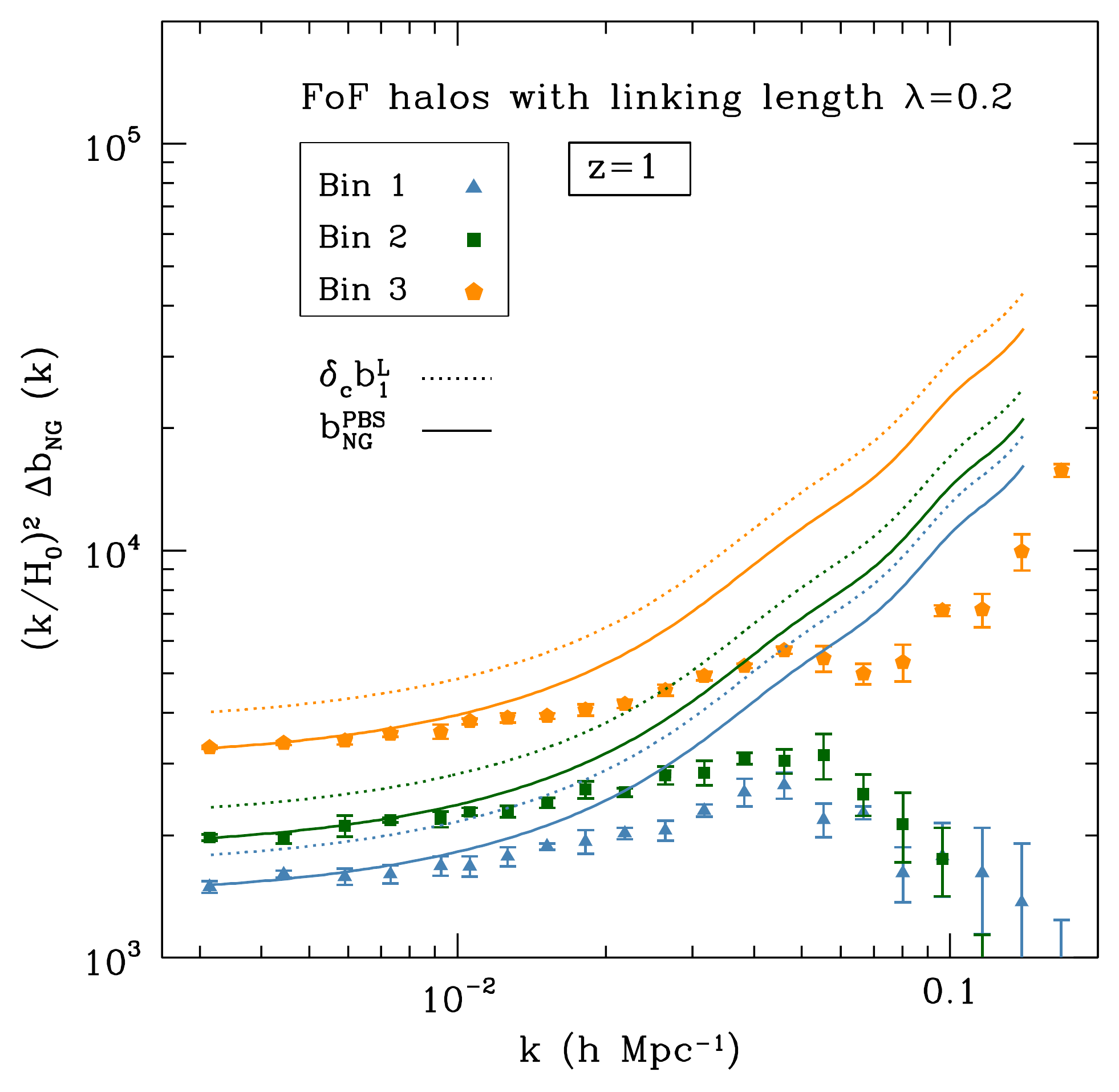}}
\resizebox{0.43\textwidth}{!}{\includegraphics{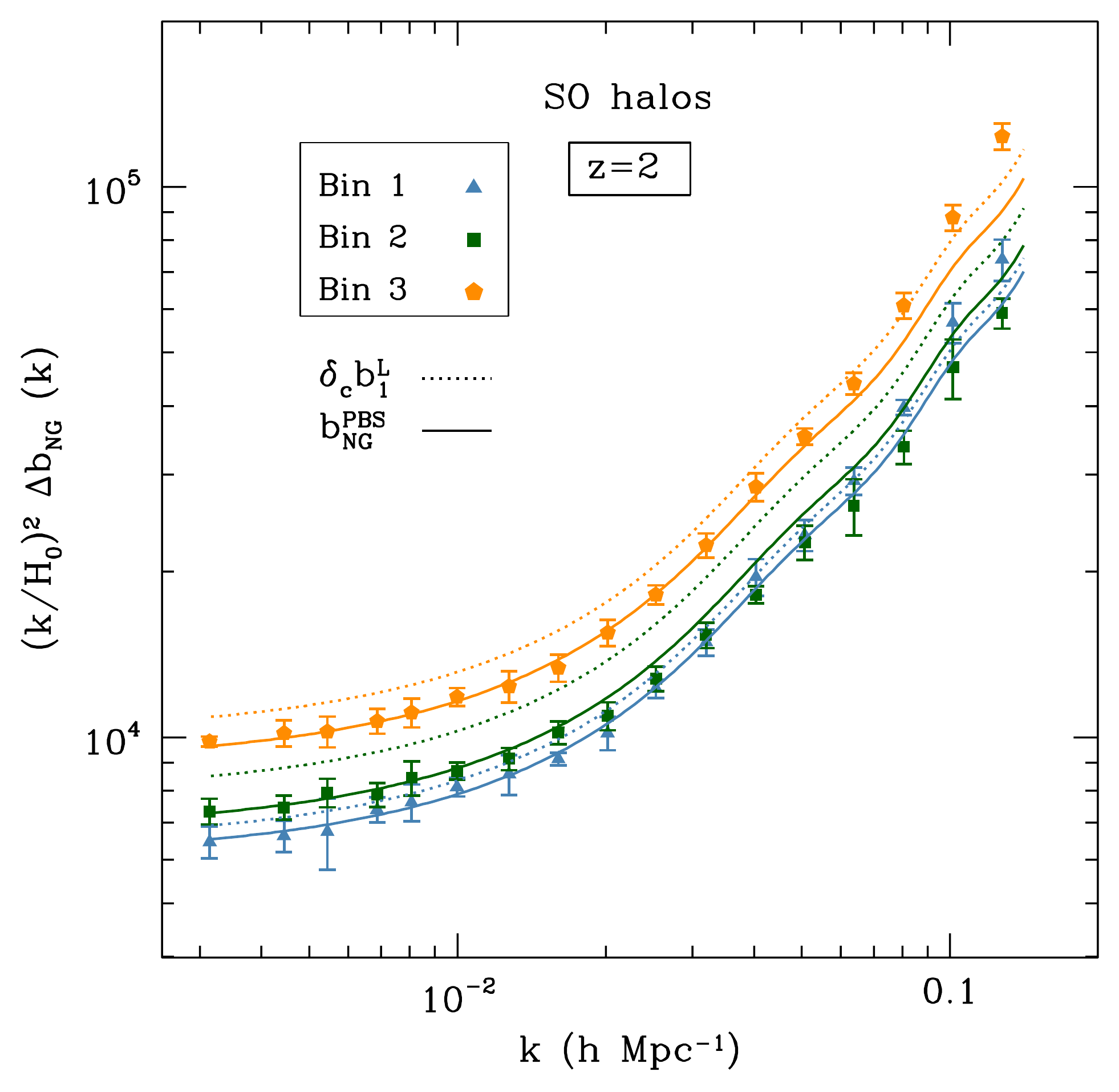}}
\resizebox{0.43\textwidth}{!}{\includegraphics{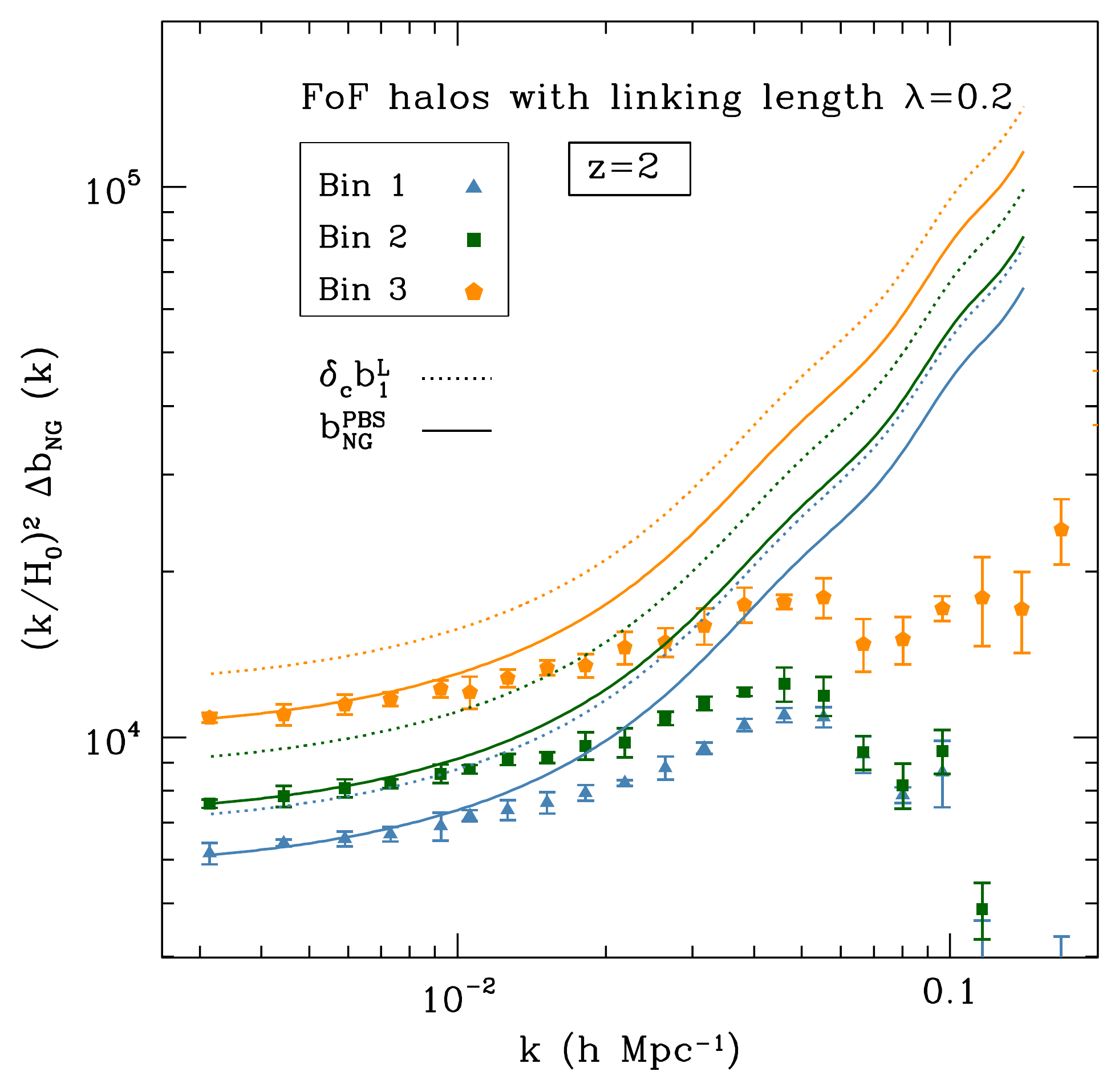}}
\caption{Non gaussian bias for the three mass bins, SO and FoF halo finder algorithms and redshifts $z=0,1,2$ for the $2$Gpc/h box sets for the non Gaussian 
simulation with $\fnl=250$. }
\label{fig:ngbiasf}
\end{figure*}

\begin{figure*}
\centering 
\resizebox{0.42\textwidth}{!}{\includegraphics{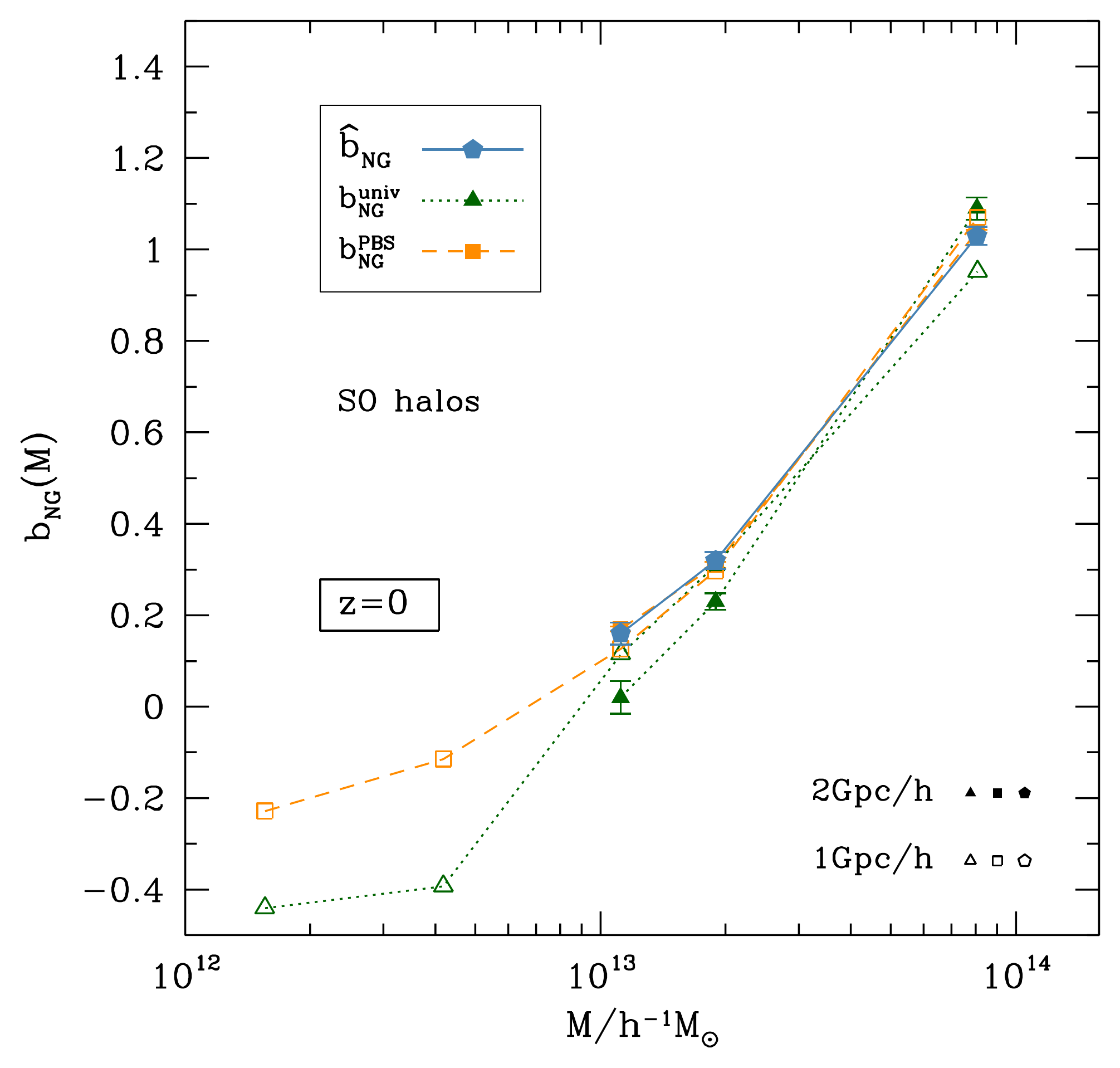}}
\resizebox{0.42\textwidth}{!}{\includegraphics{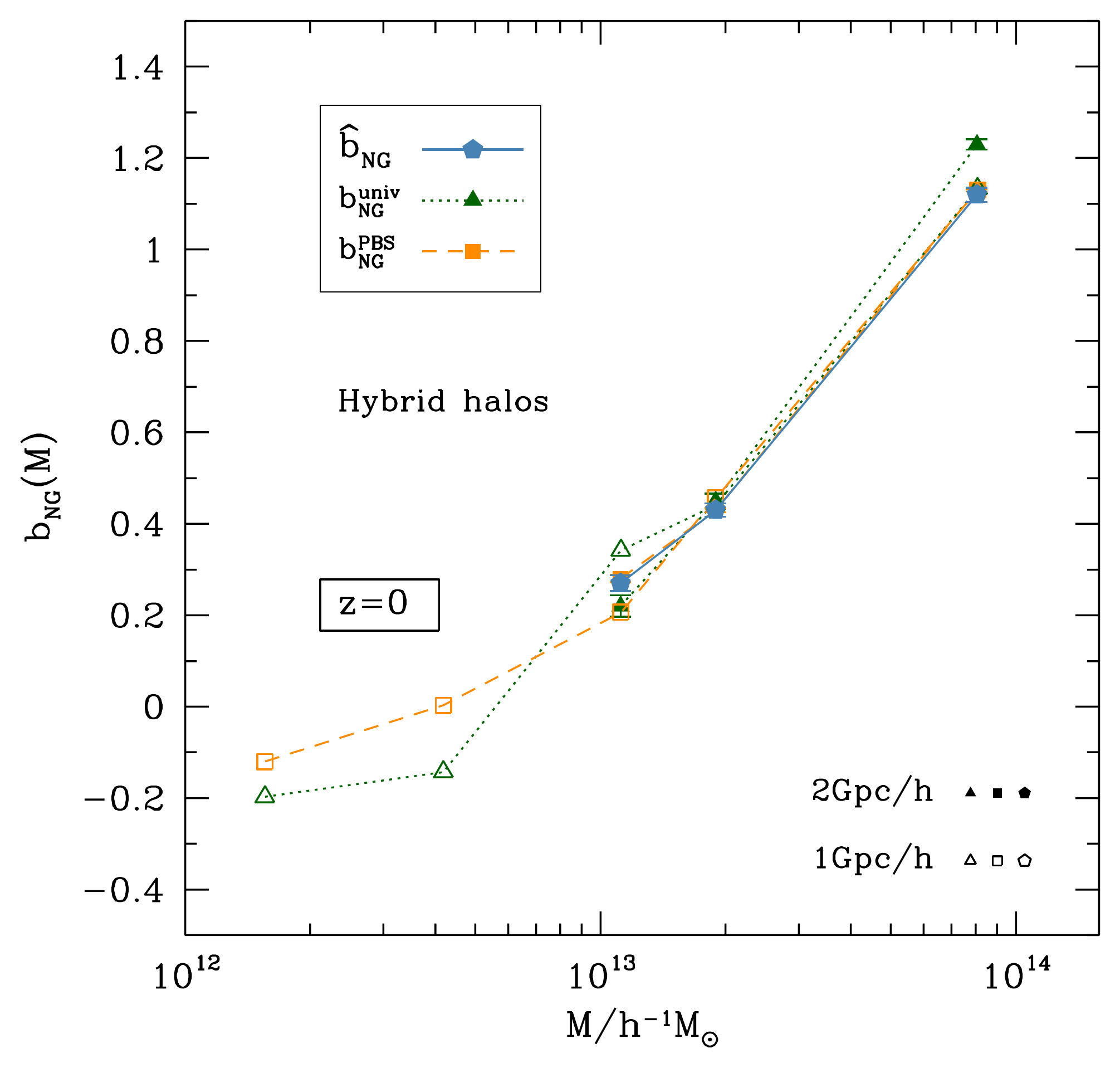}}
\resizebox{0.42\textwidth}{!}{\includegraphics{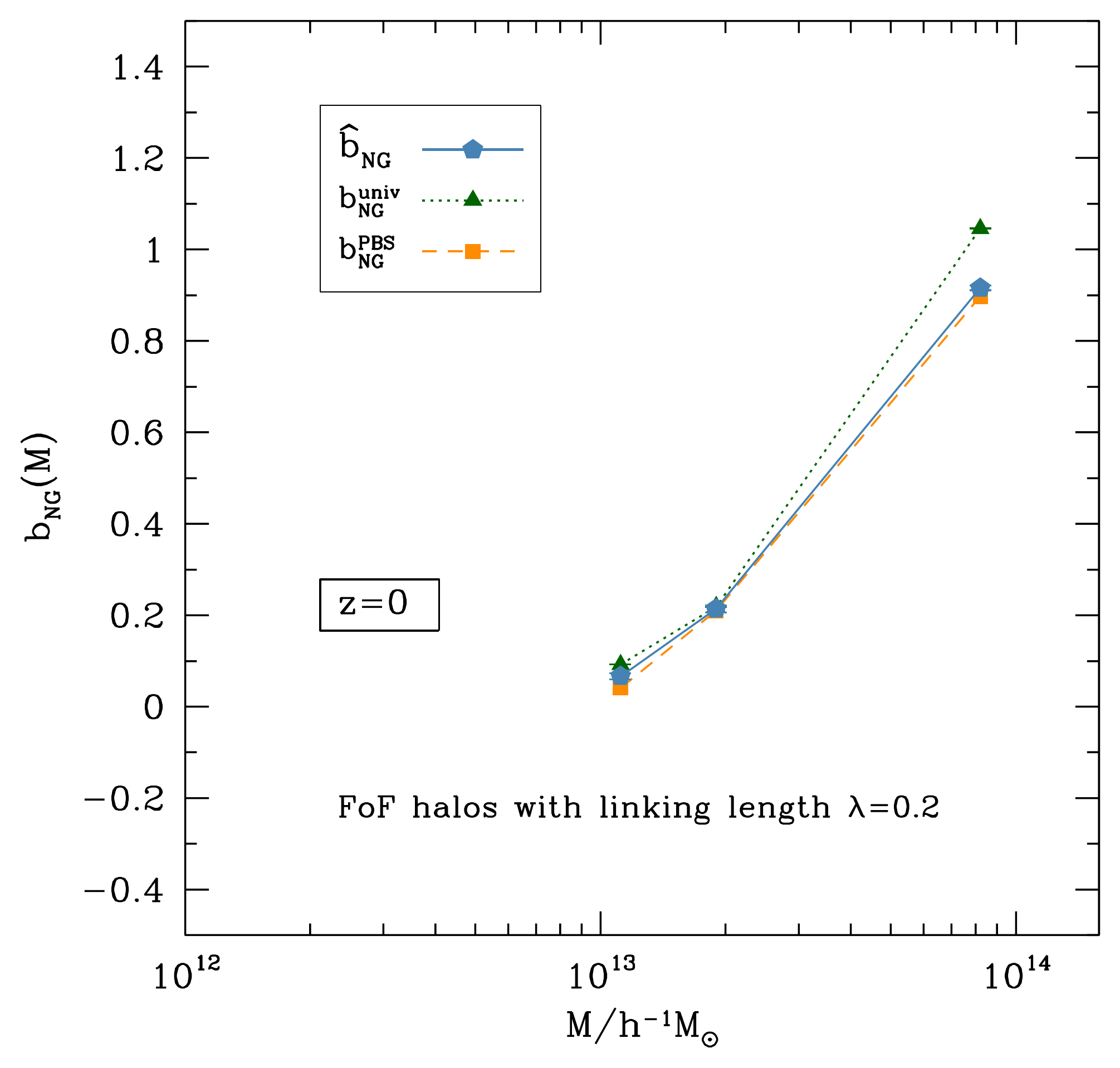}}
\resizebox{0.42\textwidth}{!}{\includegraphics{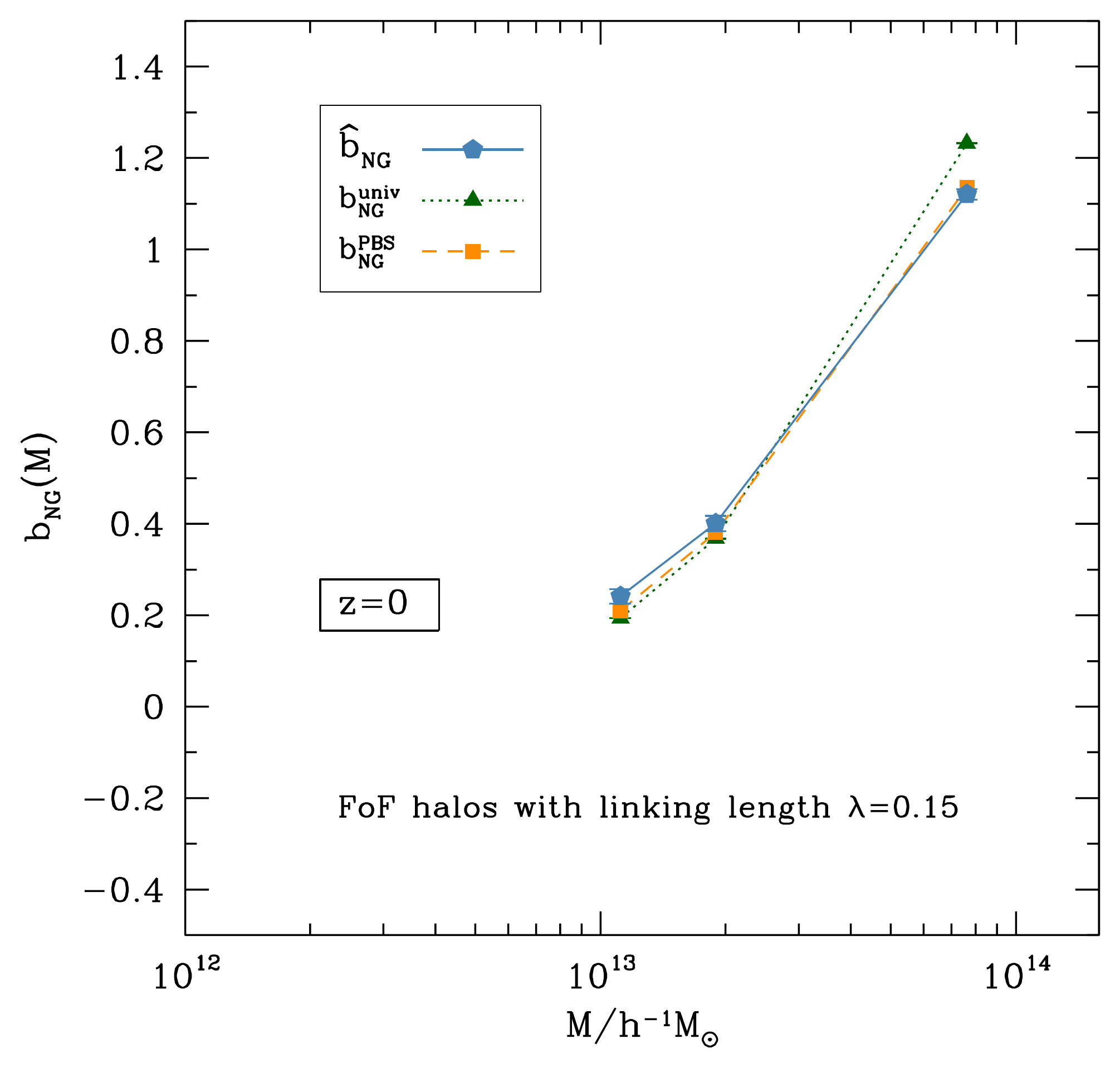}}
\caption{  Non gaussian bias amplitude $b_{\rm NG}$ as a 
function of mass for all mass bins and halo finders, at redshift $z=0$. Blue pentagons are the measured value using the relation of Eq.\eqref{eq:bhat}, 
green triangles are estimated using Eq. \eqref{eq:dalalnbody} and orange squares refer to the PBS prescription, Eq. \eqref{eq:measpbs}.}
\label{fig:bngz0}
\end{figure*}

In order to measure $\Delta b_\kappa(k,\fnl)$, we define the following quantities
\begin{align}
\mathcal{A}_+ &=\frac{P^{\rm NG}_{\rm hm}(k,+250)}{P^{\rm G}_{\rm mm}(k,0)} -b^{\rm G}_{\rm hm} - \Delta b_I(+250) - b^{\rm G}_{\rm hm} \beta_m(k,+250) \\
\mathcal{A}_- &= \frac{P^{\rm NG}_{\rm hm}(k,-250)}{P^{\rm G}_{\rm mm}(k,0)} -b^{\rm G}_{\rm hm} - \Delta b_I(-250) - b^{\rm G}_{\rm hm} \beta_m(k,-250)
\nonumber \;,
\end{align}
which are both evaluated for each of the six realizations. We therefore obtain
\begin{equation}\label{eq:a+a-}
\frac 12 \left(\mathcal{A}_+ - \mathcal{A}_-\right) =  \Delta b_\kappa(k,+ 250) + \mathcal{O}(\fnl b^{\rm G}_2, \fnl^3)\,,
\end{equation}
since $\Delta b_\kappa(k,\fnl)$ is linear in $\fnl$.  Each term in $\mathcal A_+$ and $\mathcal A_-$ can be computed directly from the simulations: the linear bias $b_{\rm hm}^{\rm G}$ is computed as explained in the previous section, the scale independent shift $\Delta b_I$ is evaluated by taking the numerical derivative of the measured halo mass function for non-Gaussian and Gaussian simulations and the matter power-spectrum correction $\beta_m$ is also estimated using the measured matter power spectra for Gaussian and non-Gaussian initial conditions. 

Using the combination in Eq. \eqref{eq:a+a-}, we are able to get rid of all the terms proportional to $\fnl^2$. Here and henceforth, we shall neglect all the contributions that depend on $b_2$ and $\fnl^3$. 

Our final estimate for the non-Gaussian bias $ \Delta b_\kappa(k,\fnl)$ is the average over the six realizations. Furthermore, we can invert Eq. 
\eqref{eq:deltabkappa} to have a measurement of the amplitude of $\Delta b_\kappa(k,\fnl)$, that is, 
\begin{equation}\label{eq:bhat}
\hat b_{\rm NG} = \frac{1}{N_{\vk}} \sum_{k_i \in [0.004,0.01]} \frac{\mathcal{M}(k_i)}{2\fnl} \Delta b_\kappa(k_i,\fnl) \;.
\end{equation}
In practice, we bin the measurements in Fourier space into equally spaced logarithmic bins of width $\Delta\log_{10} k = 0.1$, and average over all the
bins lying in the wavenumber interval $[0.004,0.01]$.\footnote{Note that the $k_{\rm max}$ we use here, $k_{\rm max} = 0.01$, is lower than the one we 
used for the Gaussian bias measurement, $k_{\rm max} = 0.03$. Since loop and higher-order bias corrections contribute at the same scales in the two cases, 
the $k_{\rm max}$ should be in principle the same. However, in the case of the non Gaussian bias there are additional uncertainties in the determination, 
for example, of the scale independent shift $\Delta b_I$, so that we use a more conservative $k_{\rm max}$.}

To ascertain the robustness of our measurement of $\hat b_{\rm NG}$, we use an additional method for the FoF halos. Namely, we consider the quantity
\begin{equation}
\begin{split}
Q=&\frac{P_{\text{hm},+250}(k)-P_{\text{hm},-250}(k)}{2 P_{\text{mm},0}(k)}\\
=&\Delta b_\kappa(k,+250)+\Delta b_I(+250)+b^{\rm G}_{\rm hm} \beta_m(k,\fnl)
\end{split}
\end{equation} 
on large scales $k<0.02\ h\text{Mpc}^{-1}$ for each of the six realizations. On these scales the non-Gaussian corrections to the matter power spectrum 
are negligibly small. 
Using the mean $\bar Q$ and standard deviation of the mean $\Delta Q$ over the six realization, we can write down the $\chi^2$
\begin{equation}
\chi^2=\sum_{k_i}\frac{1}{\Delta Q^2(k_i)}\left(\bar Q(k_i) -\hat b_\text{NG}\frac{2\fnl}{\mathcal{M}(k_i)}-\widehat {\Delta b_I}\right)^2\; .
\end{equation}
We then proceed to find the parameters $\hat b_\text{NG}$ and $\widehat {\Delta b_I}$ that minimize the above $\chi^2$ as well as their joint covariance 
matrix. The effect of $\beta_m$, for this method, is accounted for by adding a $k^2$ component in the above fit. We have performed this check and have 
found no significant changes in the inferred constraint on $\hat b_\text{NG}$ or its error.

Since we want to test the relation Eq. \eqref{eq:euleriandb}, we distinguish between the ``universal (univ)'' and ``peak-background split (PBS)'' 
predictions for the amplitude of the non-Gaussian bias: 
\begin{eqnarray}
\label{eq:dbng}
b^{\rm univ}_{\rm NG} &=&\dc (b_1^{\rm Eul}(M)-1), \\
b^{\rm PBS}_{\rm NG} &=&\frac{\partial \ln \bar n_h}{\partial \ln \sigma_8} \;.
\label{eq:pbsng}
\end{eqnarray}
In the first relation, we subtract a factor of unity from the Eulerian halo bias in order to get the linear Lagrangian halo bias, since both are
related through \citep{Mo:1995cs}
\begin{equation}
b^{\rm Eul}_1 = 1 + b_1^{\rm L} \;.
\end{equation}
Measuring $b^{\rm univ}_{\rm NG}$ is, therefore, straightforward since we have already estimated the linear Eulerian bias in the previous section 
(cf. Table \ref{tab:bias}).
We will adopt the value $\dc=1.687$ throughout, which is motivated by spherical collapse considerations \cite{Gunn:1972sv}.
By contrast, the measurement of $b^{\rm PBS}_{\rm NG}$ requires a numerical evaluation of the derivative of the halo mass function with respect to the 
normalization amplitude $\sigma_8$. Using the 4 realizations of the 3 sets with Gaussian initial conditions with different amplitude 
$\sigma_8=0.83,0.85,0.87$ we can perform this task very precisely.  Specifically, we compute this derivative via
\begin{equation}\label{eq:measpbs}
b^{\rm PBS}_{\rm NG}(M) = \frac 14 \sum_{i=1}^4 \frac{0.85}{2\bar n^i_h(M,0.85)} \frac{n^i_h(M,0.87)-n^i_h(M,0.83)}{0.02} ,
\end{equation}
and thus obtain $b^{\rm PBS}_{\rm NG}$ as a function of 
halo  mass. The halo mass function for each realization is obtained by counting halos in 30 logarithmically spaced mass bins (see Fig. \ref{fig:ahf}). 
We then interpolate it to get a smooth function of mass. To get a prediction for the three mass bins we are considering here, we weight the 
measured values of $b^{\rm PBS}_{\rm NG}$ within each bin with the interpolated measured halo mass function. 

\begin{figure*}
\centering 
\resizebox{0.32\textwidth}{!}{\includegraphics{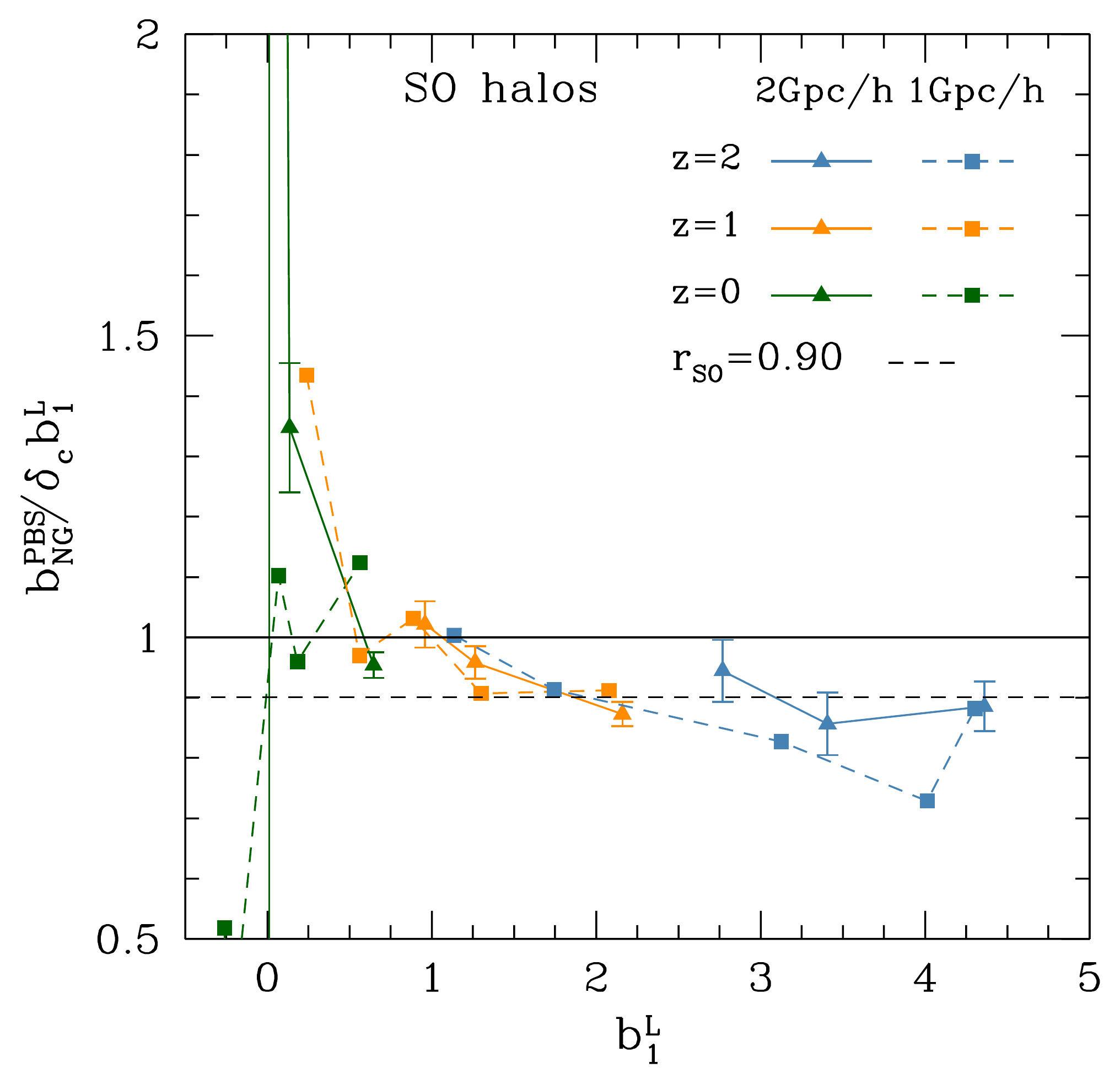}}
\resizebox{0.32\textwidth}{!}{\includegraphics{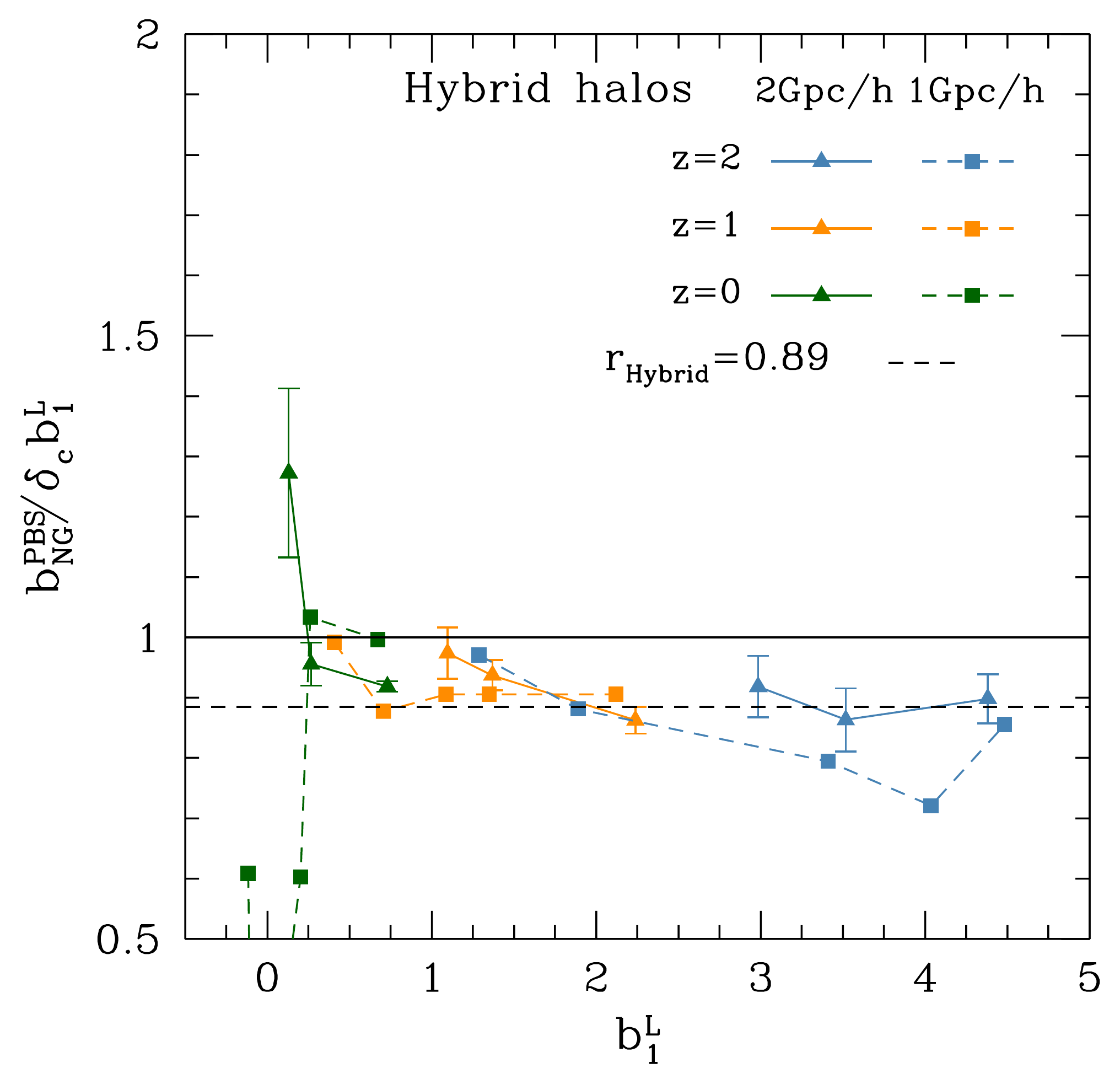}}
\resizebox{0.32\textwidth}{!}{\includegraphics{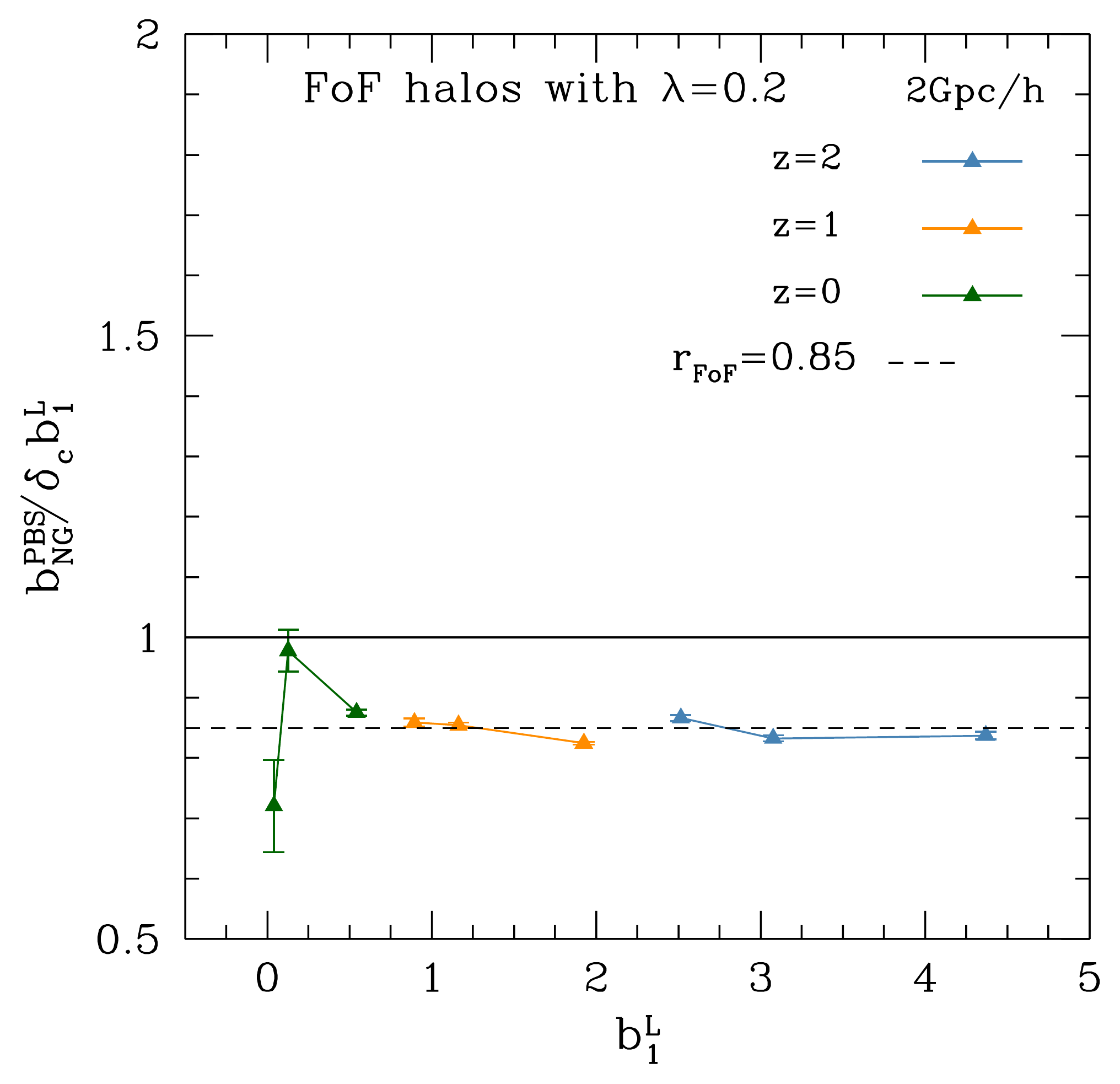}}
\caption{Ratio of the non-Gaussian bias amplitude $b_{\rm NG}^{\rm PBS}$ to the standard universal prediction $\delta_{c} b^{\rm L}_1$ as a function of $b^{\rm L}_1$ 
for all mass bins and at redshifts $z=0,1$ and $2$.  The different panels show results for the SO, Hybrid and FoF halos with linking length $\lambda=0.2$. 
Black dashed lines indicate the fitted constant value of $b_{\rm NG}^{\rm PBS}/\dc b^{\rm L}_1$ at $b^{\rm L}_1 \gtrsim1$ for each finder.}
\label{fig:bngr}
\end{figure*}

We now present the results. Firstly, we show in Figure~\ref{fig:ngbiasf} measurements of $\Delta b_\kappa (k,\fnl)$ multiplied by $(k/H_0)^2$ at redshifts $z=0$, $1$ and $2 $
for the SO and FoF halo finders. As expected from Eq.~\eqref{eq:deltabkappa}, at large scales, $k\lesssim 0.1$ h/Mpc, the scale dependence 
becomes noticeable and indeed exhibits a $k^{-2}$ behavior as is evident from the good agreement with the solid lines. 
The agreement at smaller scales, i.e. at scales $k\gtrsim 0.01$ is not as good, especially for FOF halos. This is unsurprising since we are not properly modeling those scales (i.e. including higher order terms such as higher order biases and loop corrections to the non-Gaussian power spectrum). We leave a more detailed comparison to future work.

Turning to the quantity $b_{\rm NG}$ as described in the previous paragraph, we compare the measured 
$\hat{b}_{\rm NG}$ with the estimates of $b^{\rm univ}_{\rm NG}$ and $b^{\rm PBS}_{\rm NG} $ described above. 
In Figure~\ref{fig:bngz0} we show these results at redshift $z=0$ for the three halo finders whereas, in Figure \ref{fig:bngr}, 
we include also the redshifts $z=1$ and $z=2$ for the three halo finders and plot the ratio $b_{\rm NG}^{\rm PBS}/ \dc b_1^{\rm L}$.  
If the universal prediction for $b_{\rm NG}$ were correct, then this ratio
would be equal to 1 at all redshifts.  Figure~\ref{fig:bngr} is our
main result.

\section{Discussion}
\label{sec:discussion}
We begin with Figure~\ref{fig:bngz0}.  Clearly, the peak-background split prediction $b_{\rm NG}^{\rm PBS}$, 
Eq. \eqref{eq:pbs}, is in very good agreement with
the measured scale-dependent bias $\hat b_{\rm NG}$ for all halo finders and mass 
bins considered here. As outlined in Section \ref{sec:theory}, the theory indeed predicts that $b_{\rm NG}=b_{\rm NG}^{\rm PBS}$ is exact
(at all masses). Our measurements confirm this to within errors, for the range of tested masses. 
We can therefore use our measurements of $b_{\rm NG}^{\rm PBS}$, which moreover have a smaller
statistical uncertainty,
to investigate the accuracy of the universal mass function prediction 
Eq.~\eqref{eq:dalalnbody} in detail.

Figure~\ref{fig:bngr} shows the relative deviation of $b_{\rm NG}^{\rm PBS}$, simply referred to as $b_{\rm NG}$ in the following, from the 
prediction $b_{\rm NG}^{\rm univ} = \dc b_1^{\rm L}$.  We clearly see that, for all halo finders employed,
the latter overpredicts $b_{\rm NG}$ for rare halos with $b_1^{\rm L} \gtrsim 1$.  
We have fitted the quantity
 \begin{equation}
\frac{ b_{\rm NG}^{\rm PBS}}{\dc b^{\rm L}_1}=r_{X} \qquad \mbox{for} \qquad b^{\rm L}_1 \geq 1,
\label{eq:bngrfit}
 \end{equation}
where $X$ indicates different halos finder, and find the following values,
\begin{equation}
r_{\rm SO}  = 0.9, \quad
r_{\rm Hybrid} = 0.89, \quad
r_{\rm FoF} = 0.85.
\end{equation}
For SO and hybrid halos, which show very similar behavior overall, a 
clear trend of the relative deviation with $b_1^{\rm L}$ is seen, with evidence
of a reversal for marginally biased halos with $b_1^{\rm L} \lesssim 0.5$ in
case of SO halos.  There are also strong indications that $b_{\rm NG}$ changes
sign at a nonzero value of $b_1^{\rm L}$, i.e. that $b_{\rm NG} \propto b_1^{\rm L} + $const
for these halos (see the upper left panel of Figure~\ref{fig:bngz0}). 
Further, the measurements from simulations 
with different resolution (box size) are in good agreement, with only small deviations at redshift $z=2$. Given that for the small box we have only one realization, we expect that the higher mass bins data may deviate given that the number density of halos in this range at redshift $z=2$ is low.  
Note that, since our measurements for halos at 
different redshifts have little overlap in terms of $b_1^{\rm L}$, we cannot
rule out that the quantity $b_{\rm NG}/\dc b_1^{\rm L}$ has a residual
redshift dependence in addition to that on $b_1^{\rm L}$.  
On the other hand, no significant trend with $b_1^{\rm L}$ of the deviation is 
seen for FoF halos, for which the fit in Eq.~\eqref{eq:bngrfit} is 
consistent over the entire range of $b_1^{\rm L}$.  
Our results based on the PBS expression Eq.~\eqref{eq:pbs} represent the most 
precise measurements to date of the scale-dependent halo bias due to 
primordial non-Gaussianity.

Before turning to the theoretical interpretation of our results, we briefly
compare with results in the literature.  
\cite{Pillepich:2008ka} presented simulation measurements
of $b_{\rm NG}$ for FoF halos ($\lambda=0.2$), and pointed out that 
the scale-dependent bias is smaller by 20--70\% percent than predicted by
the universal mass function.   
\cite{grossi/etal:2009} also measured the scale-dependent bias for FoF halos  ($\lambda=0.2$).  
They considered a fixed cumulative halo mass bin, $M_{\rm FoF} > 10^{13} \hmsun$, at different redshifts, 
corresponding to a range of $b_1^{\rm L} \sim 0.1 - 3$.  
Their results were found to be consistent with a uniform deviation of
\begin{equation}
\frac{b_{\rm NG}}{\dc b_1^{\rm L}} \approx 0.75\,,
\end{equation}
which they identify with an effective threshold $\tilde q \dc$ with $\tilde q=0.75$.
\footnote{We introduced the tilde to highlight the fact that \cite{grossi/etal:2009} use a rescaling $\dc\to \tilde q \dc$ 
in the ratio of non-Gaussian and Lagrangian density bias, while the S\&T99 mass function would indicate $\dc\to \sqrt{q}\dc$.}
Given their measurement uncertainties, this is most likely consistent with
our findings in the right panel of Figure~\ref{fig:bngr}.  
\cite{reid/etal:2010} analyzed the same simulations as \cite{grossi/etal:2009}, also using an FoF finder.  
Splitting halos by their formation time identified using merger trees, they find significant
dependence of $b_{\rm NG}$ on the formation time; that is, they detect assembly bias 
in the amplitude of the non-Gaussian bias.  While it would be interesting
to perform a similar study on our halo samples, we defer this to future work.  
Another analysis with FoF halos was done by \cite{Scoccimarro:2011pz}, who compare N-body measurements of the scale-dependent bias to a prediction based on the derivative of the mass function with respect to mass, and to the one predicted by universal mass functions. They find that the former is broadly consistent with the measurements, while the latter deviates to up to $50\%$, at redshift $z=0$, for halos found with a linking length $\lambda=0.2$.

\cite{Desjacques:2008vf} measured $b_\text{NG}$ using the same SO halo finder (AHF) as employed here, yet with a density 
criterion given by the redshift-dependent virial overdensity $\Delta_c(z)$ predicted by a spherical collapse 
calculation \citep{Eke:1996ds}. In particular, $\Delta_{\rm c}(z=0)\simeq 340$. 
They did not find strong evidence for the ratio $b_{\rm NG}/\dc b_1^{\rm L}$ being different from unity, although their measurements do 
not rule out a value of $b_{\rm NG}/\dc b_1^{\rm L}=0.9$ at high mass (see their Fig. 8). Another simulation set was analyzed in 
\cite{wagner/verde:2011} using the same halo definition.
Averaging over all mass and redshift bins, they obtained (their Fig.~11)
\begin{equation}
\frac{b_{\rm NG}}{\dc b_1^{\rm L}} \approx 0.9\,.
\end{equation}
They also found mild evidence for an increase in $b_{\rm NG}/\dc b_1^{\rm L}$ towards less biased 
halos, especially for their lowest two mass bins.  
These results are in very good agreement with the left panel of Figure~\ref{fig:bngr}. 

\cite{Hamaus:2011dq} presented results for both FoF and SO halos.  They derived the scale-dependent bias for a wide halo mass bin, but 
considering different weighting schemes.  Their results for FoF halos are again consistent with our results, finding a suppression of 
$b_{\rm NG}/\dc b_1^{\rm L} \approx 0.8$ with no strong mass dependence, as can be seen by comparing their results for unweighted and 
weighted halos (Figures 4 and 5 there); the latter are weighted by $b_1^{\rm L}$ to optimize the scale-dependent signal, yielding a larger 
effective halo mass and bias for the weighted halos ($b_1^{\rm L} = 0.7$ vs 0.3 for uniform weighting).  
Very different results were obtained for SO halos, for which \cite{Hamaus:2011dq} found, in the unweighted case corresponding to 
$b_1^{\rm L}=0.3$, $b_{\rm NG}/\dc b_1^{\rm L} \approx 1.4$.  This reduces in the weighted case ($b_1^{\rm L}=0.8$) to 
$b_{\rm NG}/\dc b_1^{\rm L} \approx 1.0$.  Although our measurements at very low $b_1^{\rm L}$ are not sufficiently precise to conclusively 
confirm these results for $b_1^{\rm L}=0.3$, they are broadly consistent. Moreover, our results for $b_1^{\rm L}=0.8$ indeed confirm a value of 
$b_{\rm NG}/\dc b_1^{\rm L} \approx 1$ (left panel of Figure~\ref{fig:bngr}).  

\cite{Baldauf:2015vio} also measured the response of halo counts to a rescaling of the linear power spectrum amplitude, i.e. our 
$b^{\rm PBS}_{\rm NG}$.  Further, they measured the linear bias $b_1^{\rm L}$ from the response of halo counts to a long wavelength overdensity 
implemented as an effective curvature, all for FoF haloes with linking length $\lambda=0.2$.  Combining the two measurements, they find
 that $b^{\rm PBS}_{\rm NG}/\dc b_1^{\rm L}\approx 0.85$, which is completely consistent with our findings.

Overall, we thus find good agreement with previous results on the scale-dependent bias presented in the literature.  
However, by using Eq.~\eqref{eq:pbs} we are able to provide substantially more precise constraints on $b_{\rm NG}$ for highly biased halos.  

We now discuss the implications of our results for quantitative models of halo formation.  
At high mass, discrepancies between $b_{\rm NG}^{\rm univ}$ and the data can be explained by differences in the effective spherical collapse 
threshold $\dc$, which depends on the details of the halo identification algorithm \citep[see e.g.][]{Desjacques:2010jw}.  
This could be formalized with the Lagrangian approach of \cite{Matsubara:2012nc}, which predicts a generic multiplicative factor of 
$\tilde q$ in Eq. \eqref{eq:euleriandb}.
For instance, it is known that FoF halos with linking length 0.2 trace linear overdensities with an effective linear threshold $<1.687$, 
which would explain why $b_{\rm NG}^{\rm univ}$ with $\dc=1.687$ overestimates $b_{\rm NG}$ at high mass.  
For the FoF haloes, the fitted correction factor $r_{\rm FoF}$ is consistent with the ellipsoidal collapse prediction $r_{\rm FoF}=\sqrt{q}$, 
where $q$ is the value required for the S\&T99 fit of the mass function in Fig.~\ref{fig:ahf}. The smaller linking length $\lambda=0.15$ requires 
a larger $q=0.8$ for the mass function fit and requires a consistently larger $r_{\rm FoF}$. This finding is in line with the interpretation that 
smaller linking lengths lead to more spherical haloes, which are thus in closer agreement with the spherical collapse predictions.
However, this effect is not expected to apply to SO halos.  Moreover, the departure from $b_{\rm NG}^{\rm univ}$ observed for SO halos at low mass 
cannot be reabsorbed by a change in the overdensity criterion used in the definition of SO halo masses (here, $\Delta=200$ with respect to 
matter).  This is because such a change would affect the results even more strongly at high mass, where the mass function is steep.
Therefore, the departure from universality observed here is unrelated to the effect discussed in \cite{Tinker:2008ff}, which is induced by their 
particular choice of $\Delta$ as recently pointed out by \cite{Despali:2015yla}.  
Another possible explanation is the failure of the spherical collapse approximation at low mass, which we have assumed to compute 
$b^{\rm univ}_{\rm NG}=\dc b_1$. 
One may be tempted to replace the critical threshold $\dc=1.687$ for spherical collapse by, e.g., the corresponding value 
$\dc=\delta_\text{ec}$ in the ellipsoidal collapse \citep[see e.g.][for instance]{Afshordi:2008ru}. 
However, this would most likely only explain part of the deviation, since we see significant evidence than $b_{\rm NG}$ changes sign at a 
different mass than that corresponding to $b_1^{\rm L}=0$, which cannot be explained by a change of $\dc$.

Furthermore, our findings also invalidate the non-Gaussian bias prediction of current excursion set peak (ESP) implementations. In this approach 
\citep[see][for details]{Paranjape:2012ks,Paranjape:2012jt}, the amplitude of the non-Gaussian bias is a weighted sum of all the second-order
bias parameters \citep[][]{Desjacques:2013qx}. This generally holds for any ``microscopic'' Lagrangian bias models \citep[][]{Matsubara:2012nc}, 
in contrast to models which perform a large-scale bias expansion \citep{PBSpaper}.
However, while the ESP predicts $b_\text{NG}=b_\text{NG}^\text{PBS}$ for a deterministic barrier, in agreement with our findings \citep[][]{Desjacques:2013qx}, 
the stochastic barrier of \cite{Paranjape:2012jt} yields $b_\text{NG} > b_\text{NG}^\text{PBS}$ \citep[][]{Biagetti:2015exa}, which is clearly ruled out 
by our measurements. To remedy this issue, one should treat the scatter around the mean barrier as a field with long-range correlations, rather than a 
pure white noise term as done in \cite{Paranjape:2012jt,Biagetti:2015exa}. 
This is the subject of ongoing investigations. 

Another intriguing finding is the different behaviour of the large-scale stochasticity of SO and FoF halos presented in Appendix~\ref{app:stoca}. In particular, SO halos show a significantly 
stronger scale dependence of the stochasticty on large scales than FoF halos.  
If confirmed in a more detailed analysis,
this raises interesting questions about the sensitivity of the halo stochasticity and its scale dependence on the halo identification algorithm.  

\section{Conclusion}
\label{sec:conclusion}

We have confirmed the general peak-background prediction, Eq.~\eqref{eq:pbs}, for the scale-dependent bias $b_{\rm NG}$ from local primordial non-Gaussianity for a range of halo finders and halo mass bins.  As this merely involves taking a derivative of the halo mass function with respect to the initial power spectrum amplitude, this provides a simple and efficient means to measure $b_{\rm NG}$.  No implementation of non-Gaussian initial conditions is needed at linear order in 
$f_{\rm NL}$.  Moreover, this technique can also be applied directly to simulations that employ, for example, hydrodynamics, cooling, star formation, and feedback, or semi-analytical approaches to generate galaxy catalogs from simulation outputs.  
Our results indicate that the dependence of $b_{\rm NG}$ on the linear Lagrangian halo bias $b_1^{\rm L} = b_1-1$ is typically suppressed by $10-15\%$ relative to the universal mass function prediction. This raises interesting questions for theoretical models of halo formation such as the excursion set peaks approach.

\section*{Acknowledgment}

We thank Roman Scoccimarro, Ravi Sheth, Marcello Musso and Marilena LoVerde for helpful discussions. M.B. and V.D. acknowledge support by the Swiss National Science Foundation.  
M.B. also acknowledges support from Delta ITP consortium, a program of the Netherlands Organisation for Scientific Research (NWO) that is funded by the Dutch Ministry of Education, Culture and Science (OCW). T.B. acknowledges support from a Starting Grant of the European Research Council (ERC STG Grant 279617).
F.S. acknowledges support from the Marie Curie Career Integration Grant (FP7-PEOPLE-2013-CIG)
``FundPhysicsAndLSS,'' and Starting Grant (ERC-2015-STG 678652) ``GrInflaGal'' from
the European Research Council.

\bibliography{references}

\appendix

\section{The stochasticity matrix}\label{app:stoca}
\begin{figure*}
\centering 
\resizebox{0.45\textwidth}{!}{\includegraphics{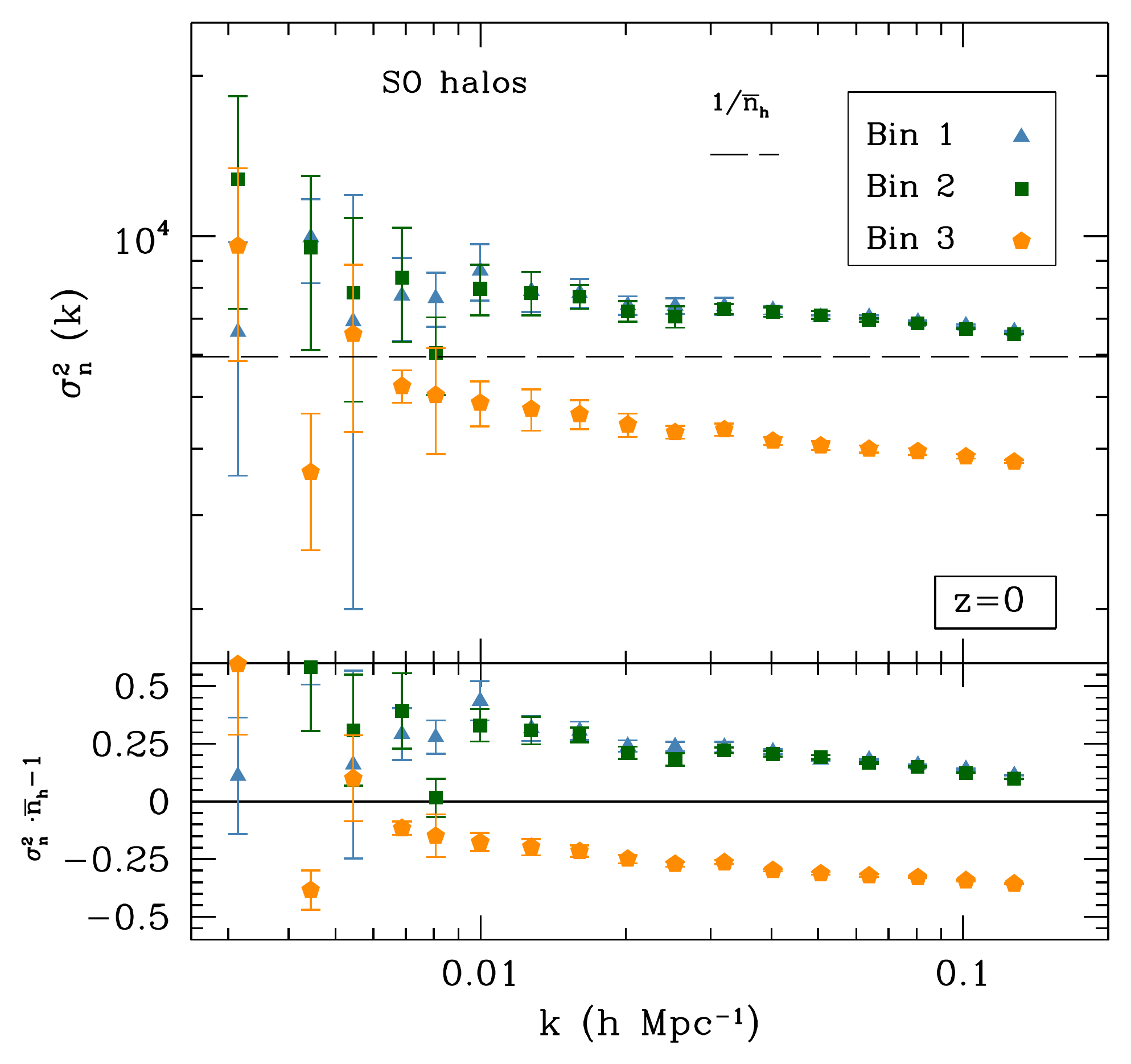}}
\resizebox{0.45\textwidth}{!}{\includegraphics{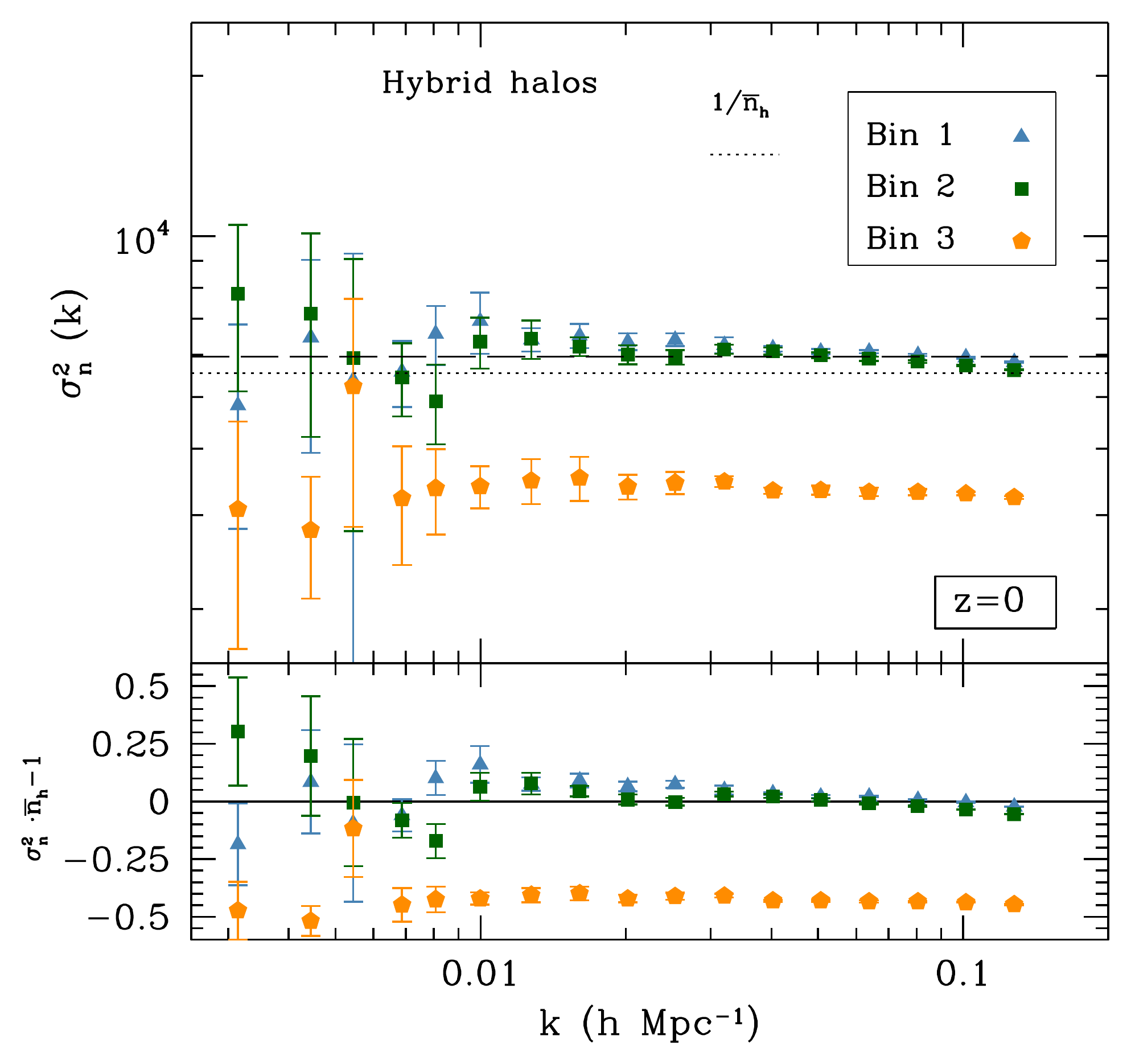}}
\caption{In the upper panel, we plot the stochasticity matrix as a function of $k$ for the three mass bins and both halo finder algorithms for the $2$Gpc/h box sets for the Gaussian simulations and the constant shot noise $1/\bar n_h$ for SO (black long dashed line) and FoF (dotted line). In the lower panel, we plot the relative difference between the stochasticity matrix $\sigma^2_n$ and the constant shot noise $1/\bar n_h$.}
\label{fig:stoca}
\end{figure*}

In this appendix, we present results on the shot noise from our measurement of the halo-halo power spectrum $P_{\rm hh}$. For this purpose, let us define the stochasticity matrix as (see \cite{Seljak:2009af})

\begin{eqnarray}
\sigma^2_n(k) &=& \langle [\delta_{\rm h}(k) - b_{\rm mh} \delta_{\rm m}(k)]^2 \rangle\\
						&=& \hat P_{\rm hh}(k) - 2 b_{\rm mh} \hat P_{\rm mh}(k) + b^2_{\rm mh} \hat P_{\rm mm}(k)
\end{eqnarray}
where the hat indicates quantities measured from simulations. We plot the stochasticity matrix $\sigma^2_n$ as a function of the wavenumber $k$ in Figure \ref{fig:stoca}. Our choice of approximately equal number density mass bins corresponds to approximately equal fiducial $1/\bar{n}_{\rm h}$ shot noise power spectrum amplitudes indicated by the horizontal lines.
Similarly to what was found in \cite{Seljak:2009af,Baldauf:2013hka} for FoF halos, the {\small Rockstar} halos show an approximately 
constant noise level in the limit $k\to 0$. For the highest mass bin the measured shot noise is lower than the fiducial shot noise, but for the two lower mass bins we don't see a significant deviation from $1/\bar{n}_{\rm h}$. However, for SO halos the behavior of $\sigma^2_n$ as a function of $k$ exhibits an unexpected scale dependence at large scales, particularly evident for the lowest mass bin. Furthermore, in the $k\to 0$ limit, the two lowest mass bins exceed the fiducial shot noise significantly, whereas the higher mass bin seems to approach $1/\bar{n}_{\rm h}$.
\cite{Baldauf:2013hka} explained the negative stochasticity corrections with small scale halo exclusion and the positive corrections with non-linear clustering (for instance from second order bias $b_2$). Both of these effects vanish on small scales ($k \gg 1/R$, where $R$ is the typical size of the halo).
The observed positive large scale stochasticity correction and scale dependence for SO haloes thus hint towards significant differences between FoF and SO haloes in either the non-linear biasing or the exclusion. In particular, a reduced exclusion scale would lead to more dominant non-linear clustering effects, which typically come with a larger typical scale $R$ and thus approach zero for lower wavenumbers. We reserve a detailed investigation of this issue for future work.
\begin{figure*}
\centering 
\resizebox{0.43\textwidth}{!}{\includegraphics{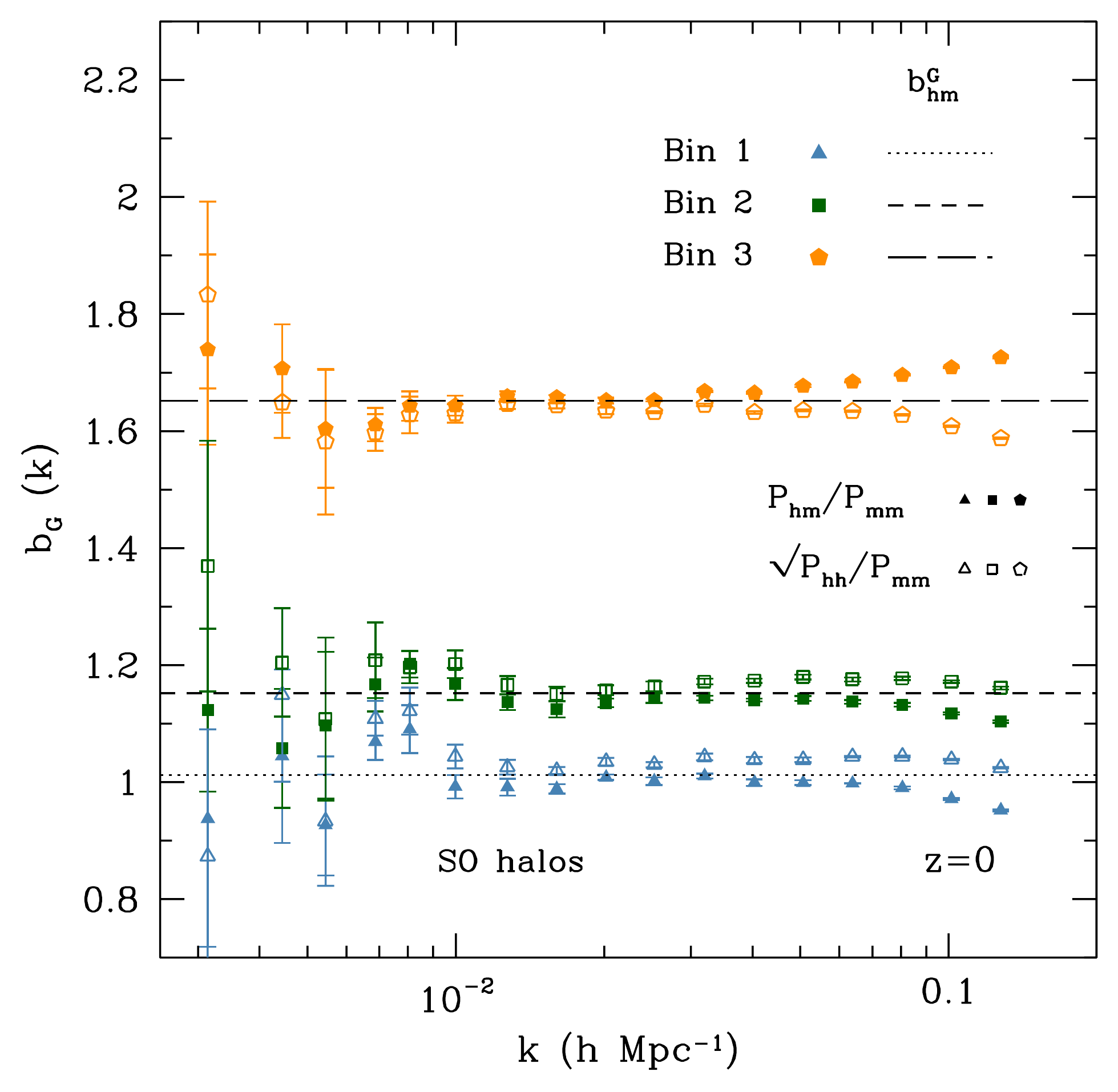}}
\resizebox{0.43\textwidth}{!}{\includegraphics{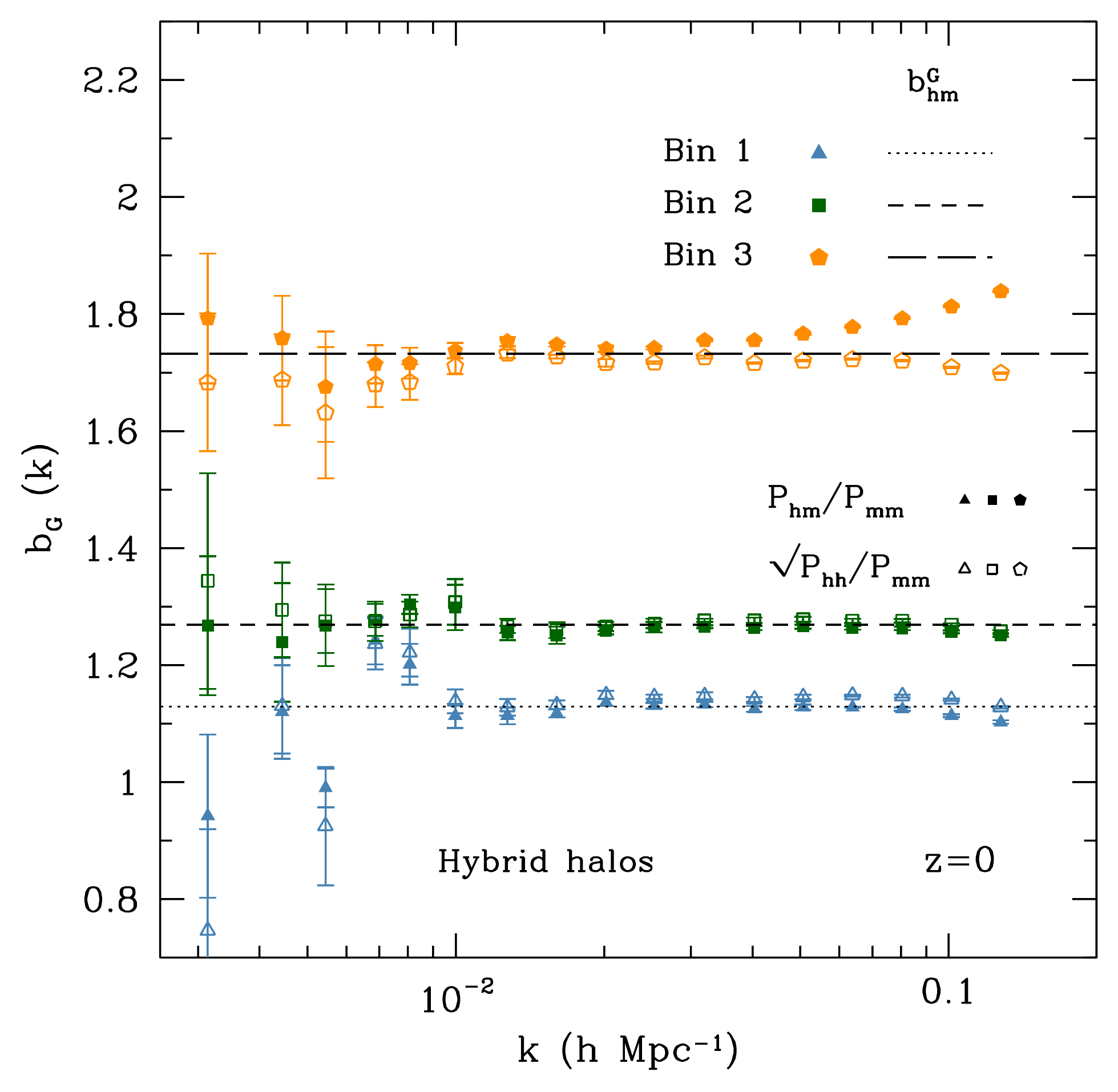}}
\resizebox{0.43\textwidth}{!}{\includegraphics{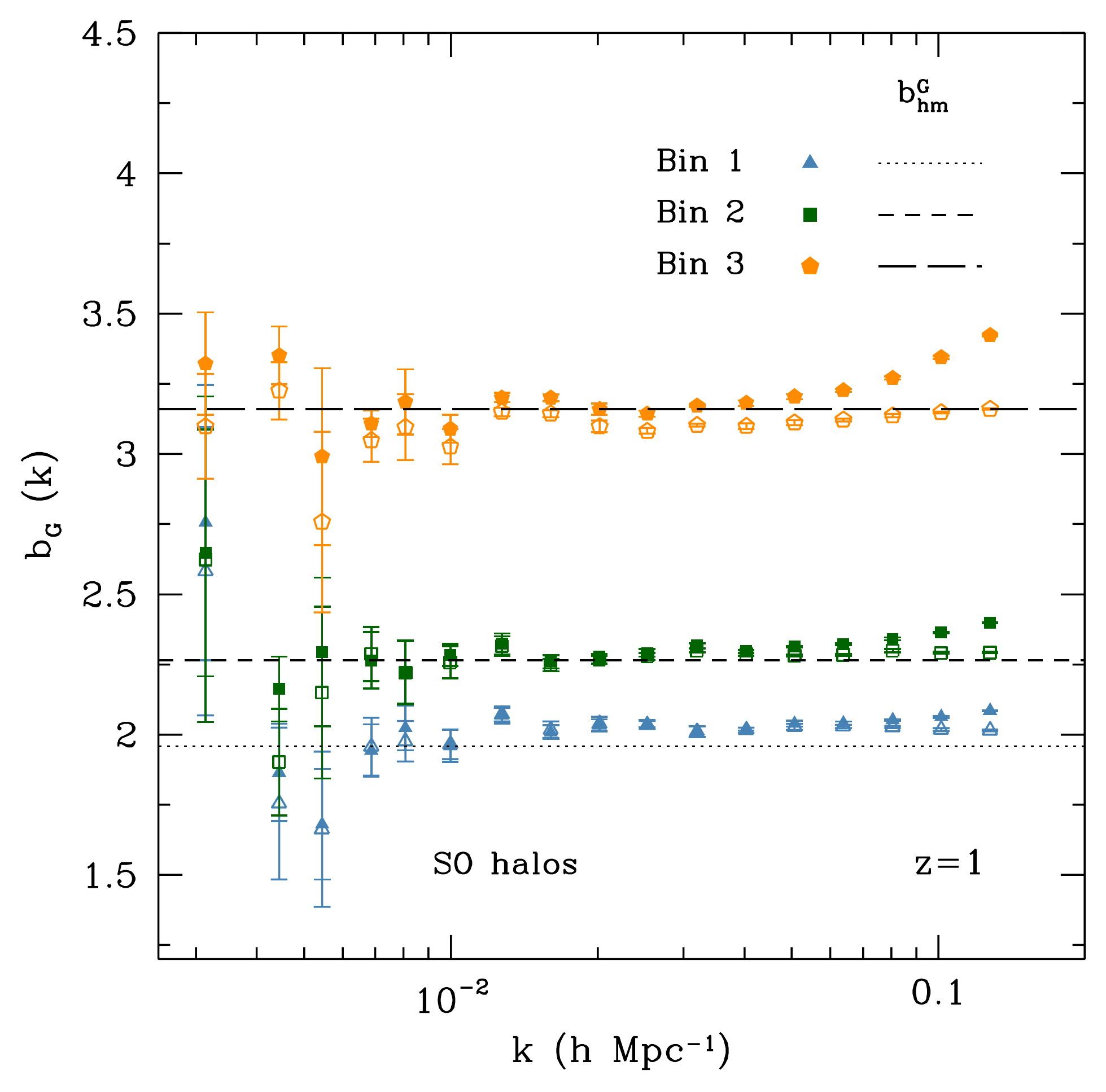}}
\resizebox{0.43\textwidth}{!}{\includegraphics{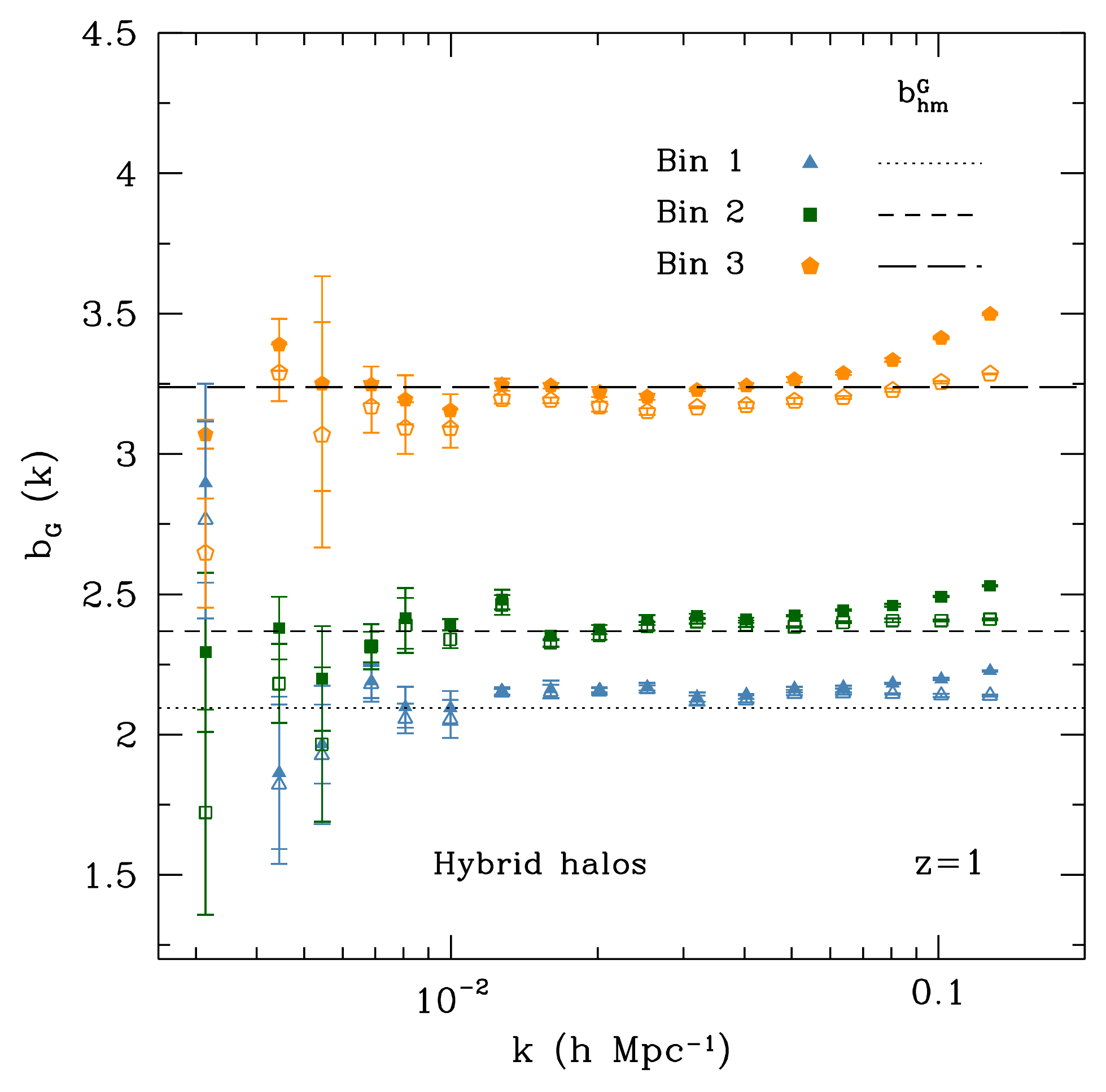}}
\resizebox{0.43\textwidth}{!}{\includegraphics{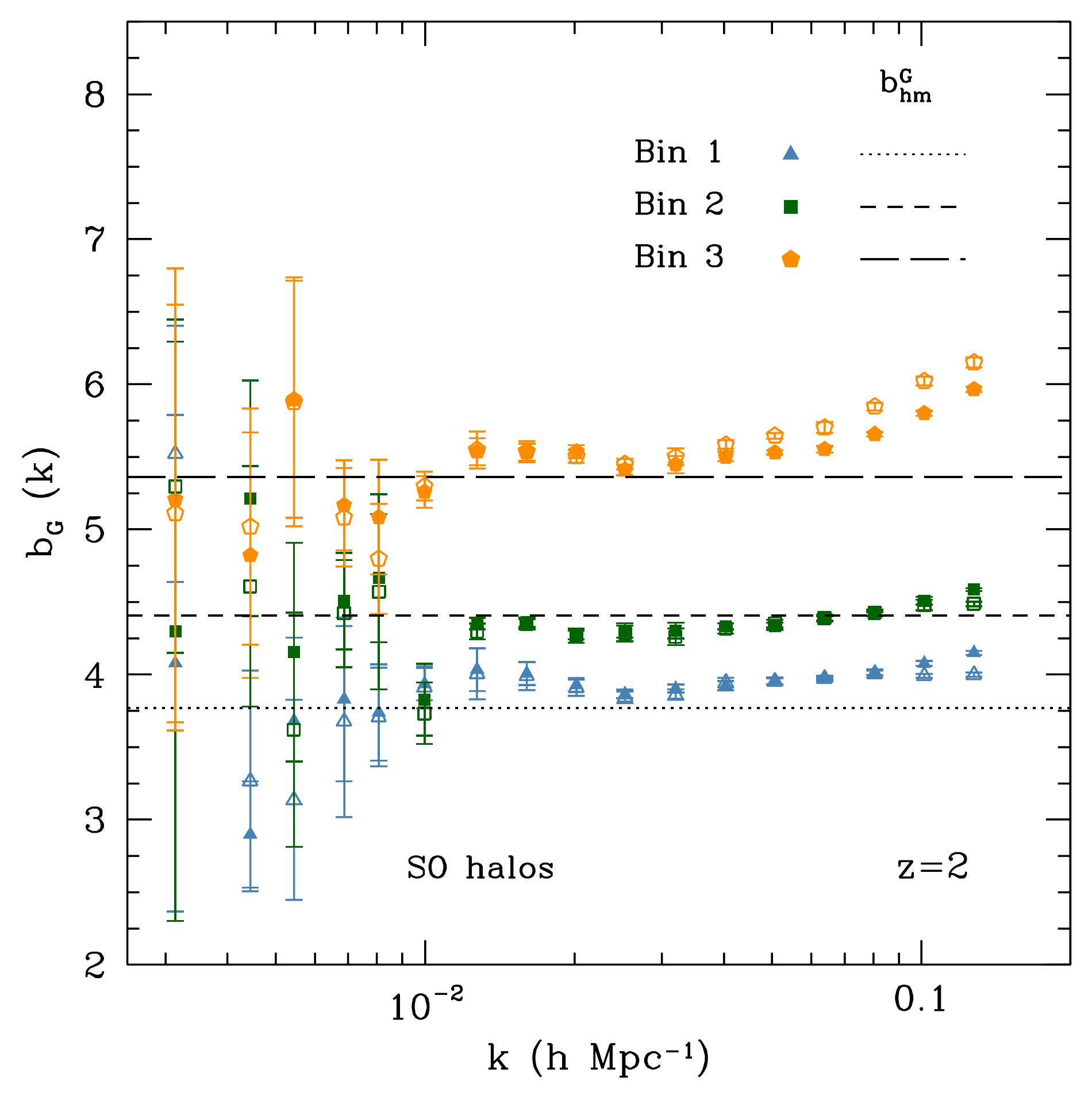}}
\resizebox{0.43\textwidth}{!}{\includegraphics{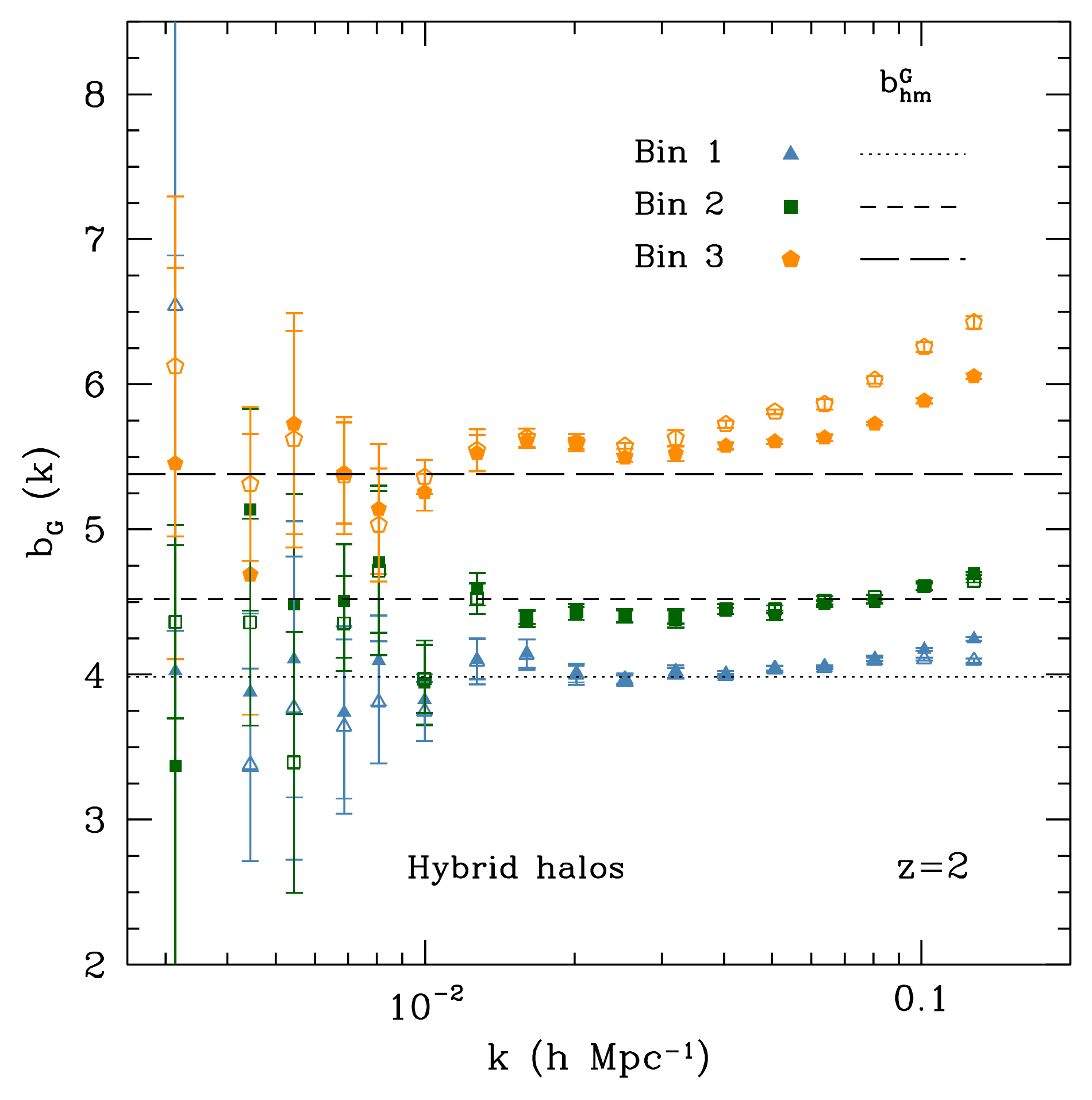}}
\caption{Linear bias for SO halos (left) and Hybrid halos (right)  at all redshifts for the Gaussian simulations with $\sigma_8=0.85$ for three mass bins, where we are using the $2$Gpc/h box sets. We have subtracted the shot-noise $1/\bar n_h$ in $P_{\rm hh}$. Horizontal lines indicate the fitted value of $b_{\rm hm}^{\rm G}$ as explained in the text.}
\label{fig:lbias}
\end{figure*}
The difference in the stochasticity is manifest also when measuring the linear bias from the halo-halo power spectrum,
\begin{equation}
b^{\rm G}_{\rm hh} = \frac{P_{\rm hh}^{\rm G}-1/\bar{n}_{\rm h}}{P^{\rm G}_{\rm mm}}.
\end{equation}
After subtracting the Poisson noise expectation from $P_{\rm hh}^{\rm G}$, the value of the linear bias $b^{\rm G}_{\rm hh}$ differs from the one inferred from the halo-matter cross power spectrum via Eq. \eqref{eq:bhm} in the case of the two lower mass bins of SO halos, even on scales where higher order biases are explected to have a negligible effect. On the other hand, the two values agree for Hybrid halos up to $k\sim 0.03$, see Fig. \ref{fig:lbias}.
%
%

\label{lastpage}

\end{document}